\documentclass[twocolumn]{emulateapj}
\begin{document}

\newcommand{\Msun}{$M_\odot$}
\newcommand{\Msunyr}{$M_\odot$yr$^{-1}$}
\newcommand{\Msunyrkpcsq}{$M_\odot$yr$^{-1}$kpc$^{-2}$}
\newcommand{\SFRSD}{$\Sigma_{\rm SFR}$} 
\newcommand{\DustSD}{$\Sigma_{\rm dust}$} 
\newcommand{\SigmaMstar}{$\Sigma_{\rm M*}$} 
\newcommand{\uJy}{$\mu$Jy}
\newcommand{\uJyperbm}{$\mu$Jy\,beam$^{-1}$} 
\newcommand{\mstar}{$M_*$}
\newcommand{\submmflux}{$S_{\rm 870\,\micron}$}
\newcommand{\ergpersec}{erg\,s$^{-1}$}

\title{ALMA 200-PARSEC RESOLUTION IMAGING OF SMOOTH COLD DUSTY DISKS IN TYPICAL $\lowercase{\it z} \sim 3$ STAR-FORMING GALAXIES}

\author{
W. Rujopakarn\altaffilmark{1,2,3},
E. Daddi\altaffilmark{4},
G. H. Rieke\altaffilmark{5},
A. Puglisi\altaffilmark{4},
M. Schramm\altaffilmark{6},
P. G. P\'erez-Gonz\'alez\altaffilmark{7},\\
G. E. Magdis\altaffilmark{8,9,10},
S. Alberts\altaffilmark{5},
F. Bournaud\altaffilmark{4},
D. Elbaz\altaffilmark{4},
M. Franco\altaffilmark{4},
L. Kawinwanichakij\altaffilmark{3},\\
K. Kohno\altaffilmark{11,12},
D. Narayanan\altaffilmark{8,13,14},
J. D. Silverman\altaffilmark{3},
T. Wang\altaffilmark{6,11,12}, and 
C. C. Williams\altaffilmark{5}
}

\affil{\phantom{$^1$Department of Physics, Faculty of Science, Chulalongkorn University, 254 Phayathai Road, Pathumwan, Bangkok 10330, Thailand\\}
$^1$Department of Physics, Faculty of Science, Chulalongkorn University, 254 Phayathai Road, Pathumwan, Bangkok 10330, Thailand\\
$^2$National Astronomical Research Institute of Thailand (Public Organization), Don Kaeo, Mae Rim, Chiang Mai 50180, Thailand\\
$^3$Kavli IPMU (WPI), UTIAS, The University of Tokyo, Kashiwa, Chiba 277-8583, Japan; wiphu.rujopakarn@ipmu.jp\\
$^4$CEA, IRFU, DAp, AIM, Universit\'e Paris-Saclay, Universit\'e de Paris,  Sorbonne Paris Cit\'e, CNRS, F-91191 Gif-sur-Yvette, France\\
$^5$Steward Observatory, The University of Arizona, Tucson, AZ 85721, USA\\
$^6$National Astronomical Observatory of Japan, Mitaka, Tokyo, 181-8588, Japan\\
$^7$Centro de Astrobiolog\'{\i}a (CAB, CSIC-INTA), Carretera de Ajalvir km 4, E-28850 Torrej\'on de Ardoz, Madrid, Spain\\
$^8$Cosmic Dawn Center, Niels Bohr Institute, University of Copenhagen, Juliane Mariesvej 30, 2100, Copenhagen, Denmark\\
$^{9}$Niels Bohr Institute, University of Copenhagen, DK-2100 Copenhagen $\O$\\
$^{10}$Institute for Astronomy, Astrophysics, Space Applications and Remote Sensing, National Observatory of Athens, 15236, Athens, Greece\\
$^{11}$Institute of Astronomy, The University of Tokyo, Osawa, Mitaka, Tokyo 181-0015, Japan\\
$^{12}$Research Center for the Early Universe, The University of Tokyo, 7-3-1 Hongo, Bunkyo-ku, Tokyo 113-0033, Japan\\
$^{13}$Department of Astronomy, University of Florida, 211 Bryant Space Sciences Center, Gainesville, FL 32611 USA\\
$^{14}$University of Florida Informatics Institute, 432 Newell Drive, CISE Bldg E251, Gainesville, FL 32611, USA
}

\begin{abstract}
We present high-fidelity, 30 milliarcsecond (200-pc) resolution ALMA rest-frame 240 \micron\ observations of cold dust emission in three typical main-sequence star-forming galaxies (SFGs) at $z \sim 3$ in the Hubble Ultra-Deep Field (HUDF). The cold dust is distributed within the smooth disk-like central regions of star formation $1 - 3$ kpc in diameter, despite their complex and disturbed rest-frame UV and optical morphologies. No dust substructures or clumps are seen down to $\simeq 1- 3$ \Msunyr\ (1$\sigma$) per 200-pc beam. No dust emission is observed at the locations of UV-emitting clumps, which lie $\simeq 2-10$ kpc from the bulk of star formation. Clumpy substructures can contribute no more than $1-7$\% of the total star formation in these galaxies (3$\sigma$ upper limits). The lack of star-forming substructures in our HUDF galaxies is to be contrasted with the multiple substructures characteristic of submillimeter-selected galaxies (SMGs) at the same cosmic epoch, particularly the far-IR-bright SMGs with similarly high-fidelity ALMA observations of \citet{Hodge19}. Individual star-forming substructures in these SMGs contain $\sim10-30$\% of their total star formation. A substructure in these SMGs is often comparably bright in the far-infrared as (or in some cases brighter than) our typical SFGs, suggesting that these SMGs originate from a class of disruptive event involving multiple objects at the scale of our HUDF galaxies. The scale of the disruptive event found in our main-sequence SFGs, characterized by the lack of star-forming substructures at our resolution and sensitivity, could be less violent, e.g., gas-rich disk instability or minor mergers.
\end{abstract}


\section{Introduction}\label{sec:intro}

A key achievement in the past two decades has been the progress in understanding the evolution of the {\it spatially-integrated} properties of galaxies across cosmic time. The cosmic histories of the star formation rate (SFR), stellar mass build up, and massive black hole accretion have been constrained out to the epoch  of reionization \citep[e.g.][]{MadauDick14, HeckmanBest14, Grazian15}. With appropriate sets of parameters, models of galaxy evolution can reproduce these histories as well as the general properties of today's galaxies   \citep[e.g.][]{Behroozi13, SomervilleDave15, Schaye15, Springel18}. However, many of the most fundamental processes are not well understood, especially down to {\it sub-galactic} scales, where pressing frontier questions in galaxy evolution lie: How did galactic spheroids form? How did galaxies and their supermassive black holes co-evolve?

There is a broad consensus that galaxies assemble most of their stellar mass via accretion of cold gas, which leads to gas-rich, unstable disks and in-situ disk-wide star formation \citep[`cold mode' accretion;][]{Noguchi99, Immeli04, Keres05, Bournaud07, Dekel09, Ceverino10, Inoue16}. Multiple lines of evidence support this consensus, such as the relationship between star formation and stellar mass at $z \sim 0 - 6$ \citep[the `main sequence', e.g.,][]{Brinchmann04, Noeske07, Salim07, Wuyts11, Whitaker12, Speagle14, Schreiber15, Salmon15} and the rarity of compact starbursts at $z \sim 2$ as indicated by the distribution of specific star-formation rates (sSFR) and of the infrared colors \citep{Rodighiero11, Elbaz11}. Observations of individual $z \sim 1 - 3$ galaxies provide further support. Spatially-resolved kinematic observations show that typical star-forming galaxies\footnote{Typical star-forming galaxies are defined here as those with SFR within a factor of four of the main-sequence \citep[e.g.,][]{Rodighiero11} at their corresponding redshift on all of the following main-sequence parameterizations: \citet{Whitaker12, Speagle14}, and \citet{Schreiber15}, hereafter `typical SFGs'} are isolated (i.e., not undergoing major mergers), rotation-dominated systems \citep[e.g.,][]{FS09, Wisnioski15, Stott16}. Dust-independent, sub-arcsecond radio continuum imaging and stacking further reveal that typical SFGs have intensely star-forming regions a few kpc in diameter  \citep{Lindroos16, Rujopakarn16}, comparable to the typical sizes of massive galaxies at the same epoch, despite forming stars at rates only achievable in the local Universe in the compact nuclei of galaxy mergers  \citep[e.g.,][]{Muxlow05, Rujopakarn11}. These observations suggest that we are on the right track toward understanding how typical massive galaxies were assembled.

A key prediction from simulations of star formation being fed by cold-mode accretion is the fragmentation of the disk and the emergence of star-forming clumps. The inward migration of these clumps is an integral part of the bulge formation scenario \citep[e.g.,][]{Bournaud07, Agertz09, Dekel09, Ceverino10, Mandelker14} and could be an intermediary regulating the bulge--SMBH relationship \citep{Martig09, GaborBournaud13}. Clumpy star formation in the formative era may also explain the bimodality in $\alpha$-abundance and metallicity ([$\alpha$/Fe] vs. [Fe/H]) in the Milky Way  \citep{Clarke19}. To play a significant role in bulge and galaxy assembly, clumps have to survive radiative and mechanical feedback long enough to accrete fresh fuel and migrate; simulations disagree whether this is possible \citep[for both sides of the argument, see, e.g.,][]{Genel12, Bournaud14}. 

When rest-frame UV images from the {\it Hubble Space Telescope} ({\it HST}) revealed irregular morphologies in $z \sim 1 - 3$ galaxies, suggesting the detection of the predicted star-forming clumps, there was a flurry of studies in the context of bulge formation \citep{Cowie95, vandenBergh96, Conselice03, ElmegreenElmegreen05}. These UV-bright star-forming clumps are ubiquitous features in $z \sim 1-3$ galaxies. They appear to be $\sim$ 1 kpc in size, contain $10^8 - 10^9$  \Msun\ of stellar mass, form stars at $1 - 30$ \Msunyr, and reside in kinematically-ordered systems. Their size, mass, SFRs, and age gradients are broadly consistent with the clump-driven bulge formation  scenario  \citep{Elmegreen07, Bournaud08, Genzel08, Elmegreen09a, Elmegreen09b, Genzel11, FS11, Livermore12, Wisnioski12, Guo12, MenendezDelmestre13, Guo15, Livermore15, Soto17, Guo18}. However, it has gradually become apparent that the UV clumps only contain $\lesssim$ $5 - 20$\% of the total star-formation in SFGs at $z \sim 2$ \citep[e.g.,][]{Wuyts12, Soto17, Guo18}.  These clumps are conspicuously absent from deep sub-arcsecond Atacama Large Millimeter/submillimeter Array (ALMA) and Karl G. Jansky Very Large Array (VLA) images of massive main-sequence SFGs. \citet{Rujopakarn16} showed that the UV-selected clumps are often peripheral to the central region of star formation, which is dust-obscured and can only be traced with extinction-independent imaging at longer wavelengths. \citet{Cibinel17} conducted sensitive CO($5-4$) observations of an archetypal clumpy galaxy UDF6462 \citep{Bournaud08} and found that  no more than 3\% of the total molecular gas in the galaxy can be in each clump, with most of the cold gas residing in the central star-forming region. That is, the UV-selected clumps do not appear to be the dominant star-forming clumps predicted by theory. Either the predicted star-forming clumps reside in the heavily obscured regions around the nuclei and close to the bulk of star formation, or our picture of clump-facilitated bulge assembly may need to be re-thought.  

The central regions that dominate the star formation in typical massive SFGs at $z  \sim 2$ are $\simeq 4-5$ kpc in diameter \citep{Lindroos16, Rujopakarn16}. Those of starburst galaxies (i.e., those with sSFR above the scatter of the main sequence) and sub/millimeter-selected galaxies (SMGs) are smaller, $\simeq 1 - 2$ kpc in diameter \citep{Simpson15, Ikarashi15}. Even for typical massive SFGs, these regions are so dusty that all but $\sim$ 1\% of the star formation is obscured in the UV \citep{Dunlop17}. For SMGs very high obscuration, $A_V \sim 100$ mag toward the center, is not uncommon \citep{Simpson17}. Probing the structure of these regions requires an extinction-independent tracer of star formation that is capable of sub-kpc resolution. Major progress in dissecting these regions is being made by: (1) sub/millimeter observations of gravitationally lensed galaxies  \citep[often SMGs, e.g.,][]{ALMAPartnership15, Swinbank15, Tamura15, Dye15}, and (2) exploiting the resolution and sensitivity of ALMA on field galaxies \citep[e.g.,][]{Iono16, Hodge16, Oteo17, Gullberg18, Tadaki18, Hodge19}. 

For different reasons, both approaches often capture SFGs far more luminous than the typical population. First, the lens selection at sub/millimeter wavalengths tends to favor luminous SFGs at $z \gtrsim 2-4$ due to the efficient lens selection at bright far-IR fluxes and the negative K-correction at higher redshifts \citep{Negrello17}. Second, directly studying unlensed galaxies requires a significant investment of ALMA time. Hence, the early efforts were made (reasonably) on some of  the most luminous  SFGs. Examples are the $\simeq 10-100$ milliarcsecond (mas) studies of sub-millimeter galaxies (SMGs) forming stars at $1300-2800$ \Msunyr\ by \citet{Iono16} and of less luminous SMGs by \citet{Hodge16, Gullberg18}; and \citet{Hodge19}. These heroic efforts at the high-resolution frontier start to venture into the regime of typical SFGs at $z \sim 3$. Yet, while these SMGs are $\gtrsim 2-5$ times brighter in the far-IR than typical SFGs, interferometric imaging at low-to-moderate signal-to-noise ratios often results in levels of noise that could be mistaken for structure in a smooth  disk, hampering the confidence in confirming or ruling out substructures \citep{Hodge16, Gullberg18}. An even larger ALMA time investment is required for the high fidelity imaging needed to confirm (or definitively rule out) the presence of clumps. The challenge is even greater to conduct such a search in typical SFGs at $z \sim 1-3$ that are where most of the stellar mass in the Universe formed. 

In this paper, we present unprecedentedly sensitive, 30 mas resolution ALMA 870 \micron\ dust continuum observations of three typical SFGs at $z \sim 3$,  selected from the Hubble Ultra-Deep Field (HUDF) and all fitting within a single ALMA primary beam. These images reveal the structure of their obscured star formation at high fidelity. We describe the observations and data reduction in Section \ref{sec:obs}, present results in Section \ref{sec:results}, and put our results in the observational and theoretical context in Section \ref{sec:discuss}. We adopt a $\Lambda$CDM cosmology with $\Omega_M$ = 0.3, $\Omega_\Lambda = 0.7$, and $H_0 = 70$ km s$^{-1}$Mpc$^{-1}$. At $z = 3$, $1''$ then corresponds to 7.702 kpc. The \citet{ChabrierIMF} initial mass function (IMF) is adopted throughout the paper.

\section{Observations and Data Reduction}\label{sec:obs}

The three galaxies in this sample were selected from the ALMA HUDF Survey at 1.3 mm \citep{Dunlop17}. This survey provides effectively an unbiased selection by stellar mass ($M_*$) at intermediate redshift, as is evident from the finding by \citet{Dunlop17} that they detect seven out of nine galaxies in the HUDF that have $M_* \geqslant 2 \times 10^{10}$ \Msun\ at $z > 2$. The native, untapered sensitivity of the survey, 29 \uJyperbm\ rms, reaches down to the SFR level of the main sequence; all but one of the 16 galaxies detected in the survey lie within the scatter of the main sequence at their corresponding redshifts \citep{Dunlop17, Elbaz18}. Although the number of sources is modest, the sample is representative of typical SFGs at $z \sim 2$ undergoing rapid assembly. In the northeast corner of the field lies a fortuitous constellation of three galaxies --- UDF1, UDF2, and UDF7 in the \citet{Dunlop17} nomenclature --- that can fit within the 17$''$ primary beam of ALMA at 345 GHz. This affords high-fidelity imaging of three typical SFGs in one 5-hr single-pointing ALMA observation, which we will describe in this section.

\subsection{ALMA Observations}\label{sec:obs_ALMA}

The ALMA observations were taken in four observing blocks during 2017 November $23-24$ as part of a Cycle 5 program \#2017.1.00001.S. We used the ALMA Band 7 receivers in the single-continuum mode, tuned to a central frequency of 343.5 GHz. Individual spectral windows (SPWs) were centered between 336.5 and 350.5 GHz; each SPW comprises 128 channels and covers 1875 MHz, resulting in 7.5 GHz of aggregate bandwidth. The single pointing is centered on UDF2: RA = $3^{\rm h}32^{\rm m}43.53^{\rm s}$, Dec =  $-27^{\rm o}$46$'$39\farcs28 (ICRS). At this frequency, the primary beam FWHM is 17\farcs4. This affords  good  sensitivity  at the  positions  of  UDF1 and UDF7, which are 7\farcs4  and 8\farcs2  from the phase center, at which  radii  the  primary beam attenuation correction factors are 0.61 and 0.54, respectively. The observations were carried out using 43 antennas in an extended configuration with baselines ranging from 92 to 8548 m with the 5$^{\rm th}$ and 80$^{\rm th}$ percentile baseline lengths of L$_5$ = 296 m and L$_{80}$ = 3366 m, respectively. Following the ALMA Technical Handbook, the maximum recoverable scale, $\theta_{\rm MRS} \approx 0.983 \lambda/L_5$ (radians), is 0\farcs60 for the array. This is considerably larger than the extent of the dust emission of the targets at 1.3 mm, $0\farcs1-0\farcs5$ (FWHM), based on earlier ALMA observations \citep{Rujopakarn16, Rujopakarn18}.  Likewise, the nominal resolution of the array, $\theta_{\rm res} \approx 0.574\lambda/L_{80}$ (radians), is 31 mas, corresponding to 240 pc at $z = 3$.

Each of the four observing blocks was 78 min in duration, with 48 min being on-source. The calibrators were: J0522$-$3627 for bandpass and flux density scale calibrations; J0348$-$2749 for phase;  and J0329$-$2357 and J0522$-$3627 for pointing. The precipitable water vapor was $0.4-0.6$ mm during the observations. In total, the observations took 5.2 hr, with 3.2 hr being on-source integration.

\subsection{ALMA Data Calibration and Imaging}\label{sec:obs_calibration}

We processed the raw visibilities using the ALMA calibration pipeline in CASA (version 5.1.1-5) and imaged the calibrated visibilities with the CASA task {\tt tclean}. We found deconvolution (i.e., application of the CLEAN algorithm) to be necessary to mitigate the sidelobes from the sources  because they are detected at peak signal-to-noise ratios as high  as $40-50\sigma$. We experimented extensively with the parameters in {\tt tclean}. The resulting images  are  insensitive to whether source masking is employed during deconvolution.  No artifacts are observed in sources near the primary beam edge. Considering their large separations from the phase center, we further experimented with the {\tt wproject}  gridder to take into account the non-coplanar baseline effect (the $w$ term), but found no significant improvement between the gridder choice of standard versus {\tt wproject} with {\tt wprojplanes} of up to 1024. The flux distribution of UDF1, near the edge of the primary beam, is virtually identical and the peak flux only differs by 1\% between the images produced with standard gridder and one with 1024 {\tt wprojplanes}. Additionally, the {\tt tclean} task converged consistently independent of the choices of {\tt niter} or {\tt threshold}. Overall, the resulting images produced from pipeline-calibrated data are of excellent quality, with no image artifacts that would indicate data or calibration issues. 

However, a closer inspection reveals that the small-scale structures within individual galaxies do depend on two areas of the imaging parameters. Firstly, the choice of weighting scheme assigned to the visibility points controls the synthesized beam shape and sensitivity of the image. In CASA, the imaging weight is implemented as the {\tt robust} parameter of the {\tt tclean} task, ranging from $-2$ (uniform weight, smaller beam, lower sensitivity) to $+2$ (natural weight, larger beam, higher sensitivity), with 0.0 to 0.5 being the commonly adopted value range. Natural weighting (or larger {\tt robust} setting) produces a larger synthesized beam, i.e., spatially broader distribution of flux for a given fiducial sky intensity distribution. In our case, the natural-weight, untapered beam is 58 $\times$ 46 mas, whereas {\tt robust = 0.5} produces a beam that is 42 $\times$ 30 mas. This difference affects image-based measurement of morphological properties, such as the decomposition of the bulge and disk components using 2D functional fitting as is commonly employed in optical studies, especially when the component of interest has a similar intrinsic size to the beam.

Another, perhaps more subtle, imaging procedure that affects source structure is the multiscale deconvolution \citep{Cornwell08, RauCornwell11}, which is necessary to deconvolve extended sources such as our targets. Multiscale deconvolution, by design, attributes flux to successive, yet discreet scales. Inevitably, the algorithm preferentially attributes flux to the adopted scales (and further influenced by the {\tt smallscalebias} parameter that can be manually tuned to give more weight to smaller scales).  In the example of UDF2,  we find that the multiscale-cleaned image employing a set of deconvolution scales of 0 (point source), 5, and 15 pixels has 10\% lower peak flux and excess flux at the 15-pixel scale, effectively broadening the central component of the source. In this particular situation, the residual image does not necessarily reflect the goodness of fit, because deep cleaning (i.e., very large {\tt niter}) can arbitrarily move residual flux into cleaned components. This affects the image-based structural parameter measurements  (e.g.,  effective radius and S\'ersic  index). Because the fiducial structure of the target is not known {\it a priori} to allow an informed choice of deconvolution scales, and because both single-scale and multi-scale images are faithful representations of the inverse Fourier transform of the interferometric visibilities (as are an infinite number of other images), this poses a dilemma as to which image is a better representation of the true morphology.

We reiterate that the CASA-produced images contain no imaging artifacts and exhibit a clean background noise image, characteristics of high-quality data and calibration. The final CASA image was made with a cell size of 6 mas (i.e., covering the 30 mas synthesized beam with five resolution elements to aid deconvolution), Briggs weighting with the {\tt robust} parameter of 0.5, {\tt MFS} spectral definition mode, and the H\"ogbom minor cycle algorithm. The primary beam attenuation correction is done with {\tt tclean} during this step. The final image has a synthesized beam FWHM of $42 \times 30$ mas, corresponding to $320 \times 230$ pc at $z = 3$ with a position angle of $77.8^{\rm o}$. The noise in source-free regions near the phase center of the image is well fit by a Gaussian with a rms of 11 \uJyperbm.

\begin{deluxetable*}{lcccccccccc}
\tablecaption{Typical Star-Forming Galaxies in the HUDF\label{tab:HUDF_sourcetable}}
\tablehead{
\colhead{ID} & \colhead{RA} & \colhead{Dec} & \colhead{$z$} & \colhead{M$_*$} & \colhead{$L_{\rm IR}$} & \colhead{SFR} & \colhead{$M_{\rm dust}$} & \colhead{M$_{\rm gas}$} & \colhead{$f_{\rm gas}$} & \colhead{SFR$_{\rm limit}$} \\
\colhead{} & \colhead{(deg)} & \colhead{(deg)} & \colhead{} & \colhead{(log\Msun)} & \colhead{(log$L_{\odot}$)} & \colhead{(log\Msun)} & \colhead{(log\Msun)} & \colhead{(log\Msun)} & \colhead{} & \colhead{(\Msunyr)}
}
\startdata
UDF1 & 53.18347  & $-27.77666$ &  2.698  &   $10.7 \pm 0.1$ & $12.59 \pm 0.01$  &  $444 \pm 5$ &  $9.03 \pm 0.02$ & $11.0 \pm 0.2$ & $0.7 \pm 0.1$ & 1.9 \\
UDF2 & 53.18137 & $-27.77758$ &  2.696   &   $10.9 \pm 0.2$  &  $12.35 \pm 0.01$  &  $257 \pm 4$   &  $8.99 \pm 0.03$ & $11.0 \pm 0.2$ & $0.5 \pm 0.1$ & 0.9 \\
UDF7 & 53.18053 & $-27.77971$ & 2.59  & $10.6 \pm 0.2$ & $12.01 \pm 0.01$  & $116 \pm  1$  &  $8.33 \pm 0.08$ & $10.3 \pm 0.3$ & $0.3 \pm 0.2$ & 2.8
\enddata
\tablecomments{$f_{\rm gas} = M_{\rm gas}/(M_{\rm gas} + M_*)$; $z_{\rm spec}$ is reported with three decimal points, two decimal point value indicates $z_{\rm phot}$. Uncertainties of parameters from far-IR SED fitting, e.g., $L_{\rm IR}$ and SFR are statistical. SFR$_{\rm limit}$ on substructures is $1\sigma$ per 200-pc beam, details in Section \ref{sec:results_smoothness}. We assume a main-sequence $M_{\rm gas}$/$M_{\rm dust}$ of 90 here.\\}
\end{deluxetable*}

\subsection{ALMA visibility-based Data Analysis}\label{sec:obs_GILDAS}

Even if we contend with the mild dependence of source structure on the choice of imaging and deconvolution methodology, an inherent limitation of the image-plane analysis is that information from baselines longer than those corresponding to the native synthesized beam (i.e., the median synthesized beam from the entire array) is not fully utilized. Therefore, we opt to carry out quantitative structural analysis in the $uv$ plane. 

We carried out the $uv$-based morphological analysis using GILDAS (version {\tt jul18a}, with a modification to fit an arbitrary number of model components). The calibrated visibilities (Section \ref{sec:obs_calibration}) were spectrally and temporally averaged, then exported from CASA and imported to GILDAS using the {\tt exportuvfits} and {\tt fits\_to\_uvt} tasks, respectively. The four spectral windows were then combined using the {\tt uv\_average} and {\tt uv\_merge} tasks. We fitted source models to the visibilities using the task {\tt uv\_fit}, fitting all components of all three galaxies simultaneously, and subtracted the models from the data. Combinations of models in a successive progression of complexity were considered, going from a point source, to circular Gaussian, to elliptical Gaussian, until the residual image no longer contained significant peaks or negative regions  $\gtrsim$ 3 $\sigma$.

We allow all parameters to be free (i.e., no parameter fixing) in the modeling. For example, free parameters of elliptical Gaussian models were the centroid, flux,  major/minor axes,  and position angle.  While this analysis is not susceptible to imaging parameters and has the potential  to utilize information from the longest baselines, the lack of {\it a priori} knowledge of the source model remains. That the final residual image contains no perceivable subtraction artifacts and has rms noise of 10 \uJyperbm, consistent with that of the source-free region near the phase center of the {\tt tclean} image from CASA, suggests that the models provide a good fit to the sources. Lastly, as CASA produces a more accurate primary beam attenuation model, we use the primary beam information from {\tt tclean} to correct the flux estimates from the $uv$ fit.

We note that no common software platform (e.g., CASA, GILDAS, AIPS, MIRIAD) currently supports fitting with the S\'ersic profile \citep[fitting with fixed exponential profile is supported by UVMULTIFIT and GILDAS; a thorough discussion of profiles supported by each software package in the $uv$-plane modeling is given by][]{MartiVidal14}. This is primarily because there is no analytical Fourier transform of the S\'ersic profile. Nevertheless, \citet{HoggLang13} have shown that linear superposition of Gaussians (``mixture-of-Gaussian'' in the Hogg \& Lang nomenclature) can accurately describe the commonly adopted brightness profile of galaxies, including the de Vaucouleurs and S\'ersic profiles \citep[][and references therein]{HoggLang13}. In effect, this allows a S\'ersic-like profile to be represented analytically and modeled in the $uv$-plane. We found that good fits were obtained with two nested Gaussians (in one case by one Gaussian and a point spread function). We verified that these fits follow closely the behavior of S\'ersic profiles over an order of magnitude of dynamic range.

\subsection{ALMA Astrometric Accuracy and Source Morphology Fidelity}\label{sec:obs_fidelity}

To study source morphologies at tens of mas and compare them to multiwavelength images, it is vital to establish that ALMA's astrometry is accurate and that the morphology of the source is robust against interferometric artifacts. The astrometric accuracy depends primarily on (1) the signal-to-noise ratio (S/N) of the source in addition to (2) the quality of phase referencing and the positional uncertainty of the phase calibrator. According to the ALMA Technical Handbook, the theoretical astrometric accuracy at a given observing frequency, maximum baseline length, and S/N, is $\Delta p$ = 60 mas  $\times$ (100 GHz/$\nu_{obs}$) $\times$ (10 km/B$_{max}$)/S/N,  which,  for  $\nu_{obs}$  of 343 GHz, B$_{max}$ of 8.5 km, and S/N $\geqslant$ 20, typical for our observations, is about 1.0 mas. This is already below the accuracy floor achievable with the standard calibration routine, which is 1.4 mas for 350 GHz observations with a baseline of 7.5 km (ALMA Technical Handbook, and references therein). Because the phase calibrator, J0348$-$2749, is tied to the International Celestial Reference Frame to within 0.1 mas (ALMA Calibrator Source Catalogue), the standard calibration noise floor dominates the absolute astrometric uncertainties. To confirm the quality of phase referencing and calibrations independently, we measured the positional offset of the `€˜check source'€™ that is observed as a part of each schedule block, J0336$-$2644, by fitting a circular Gaussian source model to the visibilities using the procedure described in Section \ref{sec:obs_GILDAS}. We found offsets from the phase center in RA and Dec of $1.07 \pm 0.05$ and $1.65 \pm 0.04$ mas, respectively, suggesting that the absolute astrometric accuracy is indeed approaching the 1.4-mas floor. This corresponds to 5\% of the synthesized beam size or about 11 pc at $z = 3$.

Systematic uncertainties of interferometric calibrations (e.g., baseline calibration) can introduce fictitious morphology in a source. We, therefore, need to establish  the source morphology fidelity, which, to first order, can be done by confirming that a point source remains point-like through the entire observing setup and calibration. To this end, we inspected the morphology of the phase calibrator, J0348$-$2749, which has previously been constrained to be point-like at $uv_{max}$ of at least 1137 k$\lambda$ (ALMA Calibrator Source Catalogue). Our longest baselines are $\approx$ 9779 k$\lambda$. Confirming the point-like nature of the phase calibrator hence indicates the angular scale above which morphological measurement is robust. Again, a $uv$-plane model fitting (Section \ref{sec:obs_GILDAS}) shows that the phase calibrator, detected at $\simeq10^4 \sigma$, is well described by a circular Gaussian with FWHM of $2.30 \pm 0.02$ mas, i.e., there is no measurable artificial broadening beyond the 2.3 mas scale. We also conducted this experiment  on  the  flux  calibrator,  J0522$-$3627,  which  is  detected at $\simeq10^5 \sigma$ and found the size to be $\lesssim 0.5$ mas, although there is a jet-like extended component at the 0.1\% flux level. Therefore, we adopt the result from the phase calibrator that the morphology is robust at angular resolutions $\geqslant$ 2.3 mas. These experiments quantified the fidelities of astrometric and morphological measurements to an angular scale that is a small fraction of the synthesized beam, i.e., to physical scales of $20 \times 10$ pc at $z = 3$. The position and morphology resolved by the beam of $320 \times 230$ pc can therefore be studied confidently.

\begin{figure*}[ht]
\figurenum{1}
\centerline{\includegraphics[width=\textwidth]{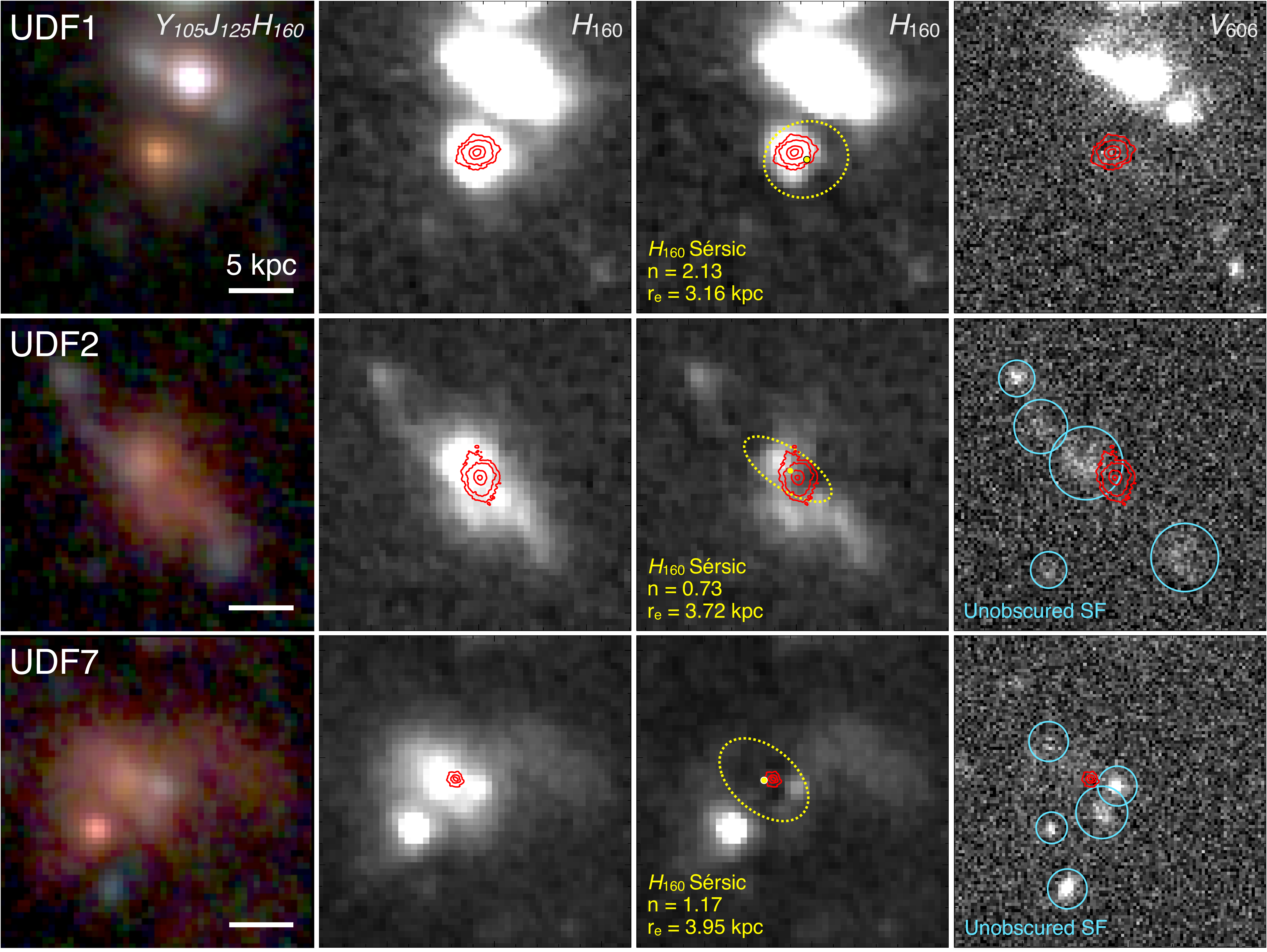}}
\caption{The locations of the sites of intense star-formation traced by ALMA (red contours) in relation to the distributions of existing stellar mass and unobscured star formation. From left to right are cutouts from (1) the {\it HST} color composites with the scale bar indicating a physical scale of 5 kpc; (2) the original {\it HST} $H_{\rm 160}$ image; (3) the {\it HST} $H_{\rm 160}$ image with the dominant S\'ersic component removed to show stellar mass substructures. The removed S\'ersic centroid and size are shown by the yellow dot and yellow ellipse, respectively (details of the GALFIT modeling to quantify this S\'ersic in Appendix A); and (4) the {\it HST} $V_{\rm 606}$ image showing rest-frame UV star formation clumps. Each cutout is $3'' \times 3''$; north is up, east is on the left. The ALMA contours are [5, 10, 25, and 40]$ \times \sigma$; the ALMA beam, not shown, is $42 \times 30$ mas, approximately the size of each pixel of the {\it HST} $V_{\rm 606}$ images in the rightmost column. The intensely star-forming regions are embedded near the centers of the stellar mass distribution, yet no stellar mass substructures are observed at their locations. The rest-frame UV morphologies of UDF2 and 7 are highly disrupted, with the unobscured star-forming clumps being spatially dislocated by $\simeq 2-10$ kpc from the obscured star-forming disk at the centers.
\label{fig:compare_HST_ALMA}}
\end{figure*}

\subsection{Ancillary Data}\label{sec:obs_ancdata}

The HUDF \citep{Beckwith06} has a uniquely sensitive set of multiwavelength imaging and spectroscopy. The following is the list of surveys and catalogs used in this work. The {\it HST} images at $0.4-1.6$ $\mu$m that reach $29.8-30.3$ mag (5$\sigma$, AB) are from the HUDF12 and XDF data releases \citep{Ellis13, Koekemoer13, Illingworth13}. We have corrected all {\it HST} images for the astrometric offsets reported by \citet{Rujopakarn16}, derived by comparing the {\it HST} and VLA positions ($\Delta$RA$ = -80 \pm 110$ mas, $\Delta$Dec$ = 260 \pm 130$ mas bring {\it HST} astrometry to an agreement with the ICRF), which is now confirmed by comparing {\it HST} positions with those from Pan-STARRS (M. Franco et al., in preparation). {\it Spitzer} and {\it Herschel} catalogs at $3.6-500$ $\mu$m are based on the methodology originally discussed in \citet{Elbaz11}; a new, deblended far-IR photometric catalog described by \citet{Elbaz18} is used in this work. The field has been a subject of intense ALMA contiguous deep-field observations, including, e.g., at 1.3 mm \citep{Dunlop17} and the ALMA Spectroscopic Survey in the HUDF (APECS; PI: F. Walter) at Bands 3 ($84-115$ GHz) and 6 ($212-272$ GHz). Ultra-deep VLA observations at 6  GHz reach 0.32 \uJyperbm\ rms \citep{Rujopakarn16},  detecting all the confirmed\footnote{One of the \citet{Dunlop17} ALMA sources, UDF14, at $3.7\sigma$ has no radio detection and has not been confirmed in more sensitive ALMA observations.} \citet{Dunlop17} ALMA sources at $z < 4$. Spectroscopic redshifts are available for UDF1 and UDF2; an accurate photometric redshift utilizing all available optical and near-infrared imaging is adopted for UDF7. The {\it Chandra} 7 Ms X-ray survey \citep{Luo17} is used to identify X-ray AGNs.

We estimate the spatially-integrated stellar mass ($M_*$) using {\tt HYPERZ} \citep{Bolzonella00} and the \citet{PPG08} SED fitting codes, utilizing the most recent revision of the \citet{Barro11} photometric compilation. All photometry out to 8.0 $\mu$m is utilized, adopting the \citet{BC03} stellar population synthesis model (hereafter BC03), experimenting with both the constant and exponentially declining star-formation histories, a \citet{ChabrierIMF} IMF, and the \citet{Calzetti00} extinction law with $A_V$ up to 5 mag. The stellar mass estimates from the two codes agreed within $0.1-0.2$ dex; we average them into the adopted values tabulated in Table \ref{tab:HUDF_sourcetable}. We further find good agreement between these values and those estimated with EAZY \citep{Brammer08} using the \citet{Conroy09} Flexible Stellar Population Synthesis, indicating well constrained stellar masses for these galaxies. We use GALFIT \citep{Peng10} to model and decompose the {\it HST} $H_{160}$ images \citep{Illingworth13} to estimate the effective radius, $r_e$, and the S\'ersic index, $n$. We fitted all components simultaneously to achieve a uniform residual noise, then subtracted the dominant S\'ersic component from the fit to search for any stellar mass substructures. The procedure is illustrated in Appendix A.

We estimate total IR luminosities, $L_{\rm IR}$, and dust masses, $M_{\rm dust}$, following \citet{Magdis12} by fitting the {\it Spitzer}, {\it Herschel} (deblended), and ALMA photometry at $24-1300$ $\mu$m with the \citet{DraineLi07} models. The total gas mass, $M_{\rm gas}$, for each galaxy, which incorporates both the molecular and atomic phases, is then inferred from its dust mass and the metallicity-dependent dust-to-gas ratio, GDR($Z$), conversion factor presented in Magdis et al. (2012). For our sample, we adopt a solar metallicity that corresponds to GDR($Z_{\odot}$) $\approx 90$. The $M_{\rm gas}$ estimates assuming GDR($Z_{\odot}$) agree within $\simeq 0.1-0.2$ dex with those inferred based on the recipe presented in \citet{Scoville17} using the monochromatic flux densities at 1.3\,mm from \citet{Dunlop17} because 870 $\mu$m no longer firmly lies in the rest-frame Rayleigh--Jeans regime at $z \geqslant 2.5$. The $L_{\rm IR}$ is used to estimate the SFR via the \citet{Kennicutt98} relation, with a factor of 0.66 adjustment to the \citet{ChabrierIMF} IMF. We note that the $L_{\rm IR}$ estimates from the \citet{DraineLi07} models agree within $\lesssim0.1$ dex with those from the \citet{Rieke09} templates. Physical parameters of HUDF galaxies are tabulated in Table \ref{tab:HUDF_sourcetable}; CANDELS ID are from \citet{Guo13}; object-specific notes on the ancillary data are as follows. 

UDF1 (CANDELS ID: 15669) has $z_{\rm spec} = 2.698$ from the ASPECS Band 3 survey \citep{GonzalezLopez19, Aravena19, Decarli19, Boogaard19}. It harbors an X-ray AGN with  $L_{\rm X,\,0.5-7\,kev} = 6.4 \times 10^{43}$ erg s$^{-1}$ (intrinsic, absorption-corrected). Its radio luminosity of $L_{\rm 6\,GHz} = 3.8 \pm 0.2 \times 10^{23}$ W Hz$^{-1}$ is consistent with being of star-forming origin \citep{Rieke09}. UDF1 is a borderline starburst, with SFR/SFR$_{\rm MS}$ of $3.1-3.5$ (SFR$_{\rm MS}$ being the SFR centered on the main sequence given its stellar mass and redshift) depending on the choice of SFR$_{\rm MS}$ parameterizations among \citet{Whitaker12, Speagle14}, and \citet{Schreiber15}.

UDF2 (CANDELS ID: 15639) has $z_{\rm spec} = 2.696$, also from the ASPECS. This galaxy is not detected in the 7 Ms {\it Chandra} X-ray observations; its radio luminosity of $L_{\rm 6\,GHz} = 3.1 \pm 0.3 \times 10^{23}$ W Hz$^{-1}$ is consistent with being of star-forming origin. UDF2 has SFR/SFR$_{\rm MS}$ of $1.3-1.6$, i.e., within the scatter of the main sequence regardless of the choice of parameterization.

UDF7 (CANDELS ID: 15381) has $z_{\rm phot} = 2.59$. It is X-ray detected with an X-ray luminosity of $L_{\rm X,\,0.5-7\,kev} = 3.7 \times 10^{42}$ erg s$^{-1}$. The radio luminosity of $L_{\rm 6\,GHz} = 7.6 \pm 0.2 \times 10^{23}$ W Hz$^{-1}$ is $\sim$ 5$\times$ the level predicted by the far-IR emission and the far-IR/radio correlation, indicating a radio AGN \citep{Rujopakarn18}. Its SFR/SFR$_{\rm MS}$ is $0.9-1.2$, again, well within the main sequence regardless of the choice of main-sequence parameterization.

 \section{Results}\label{sec:results}

Our ALMA 870 $\mu$m observations probe the rest-frame emission at $\simeq 240$ $\mu$m. No cold dust feature is found at the position of the UV-bright star-forming clumps, which are spatially distanced from the bulk of star formation near the centers (Figure \ref{fig:compare_HST_ALMA}). The central regions of intense, obscured star formation in these galaxies are resolved into up to 70 resolution elements (Figure \ref{fig:GILDAS_HUDF}). They are well modeled in the $uv$ plane with  two concentric elliptical Gaussians. We find the distributions of cold dust emission at this physical resolution to be remarkably smooth, showing no sign of clumpy star formation.

\subsection{Comparison of ALMA \& Optical Morphology}\label{sec:results_HST_compare}

Figure \ref{fig:compare_HST_ALMA} shows ALMA and optical morphologies of the targets. The best-fit model of the cold dust morphology for each galaxy is shown in Figure \ref{fig:GILDAS_HUDF} with parameters tabulated in Table \ref{tab:Comp_UDF}; the S\'ersic parameters of the dominating component of the $H_{160}$ morphology are listed in Figure \ref{fig:compare_HST_ALMA}. Rest-frame UV clumps are clearly visible in UDF2 and 7 in their {\it HST} $V_{606}$ and $i_{775}$ images. The rest-frame optical features in the {\it HST} $Y_{105}$, $J_{125}$, and $H_{160}$ images indicate considerable underlying stellar mass. Yet, in all three galaxies, the unobscured star formation traces less than $1$\% of the total SFR \citep{Dunlop17}.  We note that in comparing the optical and dust distributions, GILDAS measures the ALMA sizes as Gaussian FWHM, $\theta$, whereas GALFIT measures the {\it HST} sizes as S\'ersic effective radius, $r_e$. The two conventions are related by $\theta \sim 2.430 r_e$ \citep{Murphy17}. Morphologies of the individual sources are discussed in detail below.

\subsubsection{UDF1}

Figure \ref{fig:GILDAS_HUDF} shows the rest-frame 240 $\mu$m emission of UDF1. The cold dust emission is spatially resolved into $\simeq 60$ resolution elements (the area enclosed within the FWHM extent). It is best-modeled in the $uv$ plane by two concentric elliptical Gaussians with integrated fluxes of 1.56 $\pm$ 0.16 mJy and 1.84 $\pm$ 0.16 mJy (corrected for the primary beam attenuation), and with FWHMs of $0.67 \pm 0.04 \times 0.63 \pm 0.03$ kpc and $1.90 \pm 0.20 \times 1.76 \pm 0.18$ kpc. A single elliptical Gaussian model can only recover 83\% of the integrated flux, leaving an extended halo-like residual and necessitating a second, larger component (we will touch on the possibility of the model being a single S\'ersic in Section \ref{sec:discuss_distribution}). After removing these two components, the background noise is uniform; no residual $\gtrsim$ 3$\sigma$ remains (the local $\sigma  = 15.7$ \uJyperbm). 

The {\it HST} $H_{160}$ emission is dominated by a point source (i.e.,  a  point-spread  function)  and  a  fainter  S\'ersic  component, likely representing the AGN and disk-like emission, respectively.  The S\'ersic component has an effective radius $r_e$ = $3.16 \pm  0.17$ kpc and a S\'ersic index $n = 2.13  \pm 0.13$. There  appears to be  an offset  of 0\farcs13 between  the S\'ersic component and the AGN  (Figure 1 and Appendix A), which may be a sign of a recent interaction. The cold dust is co-spatial with the AGN/point source, That is, the dust emission is co-spatial with the AGN and offset from the stellar-mass centroid. There are no off-center star-forming clumps in the {\it HST} rest-frame UV images.

\subsubsection{UDF2}

The rest-frame 240 $\mu$m emission of UDF2 is spatially resolved into $\simeq 70$ resolution elements.  The source is well modeled with two concentric elliptical Gaussians: one bright, extended component, and a fainter compact one, nicknamed hereafter as the ``disk'' and ``core'', respectively. The disk is best-modeled by an elliptical Gaussian with a FWHM of $3.63 \pm 0.17 \times 1.51\pm 0.09$ kpc, with an integrated flux of $2.0 \pm 0.1$ mJy. The core is modeled by an elliptical Gaussian with a FWHM of $0.79 \pm 0.05 \times 0.36 \pm  0.04$ kpc and a flux of $0.80 \pm  0.05$  mJy. These two components were fitted simultaneously, yielding a combined flux of $2.80 \pm 0.1$ mJy. After subtracting the core and disk from the image, no residual peak $\gtrsim$ 3$\sigma$ remains (local $\sigma = 9.6$ \uJyperbm).

The $H_{160}$ morphology is significantly disturbed. Full morphological modeling using GALFIT requires fitting simultaneously six components to yield a uniform residual image (model shown in Appendix A). There are two dominant ones: a S\'ersic disk and a compact component indicating a substantial concentration of stellar mass. The dominant S\'ersic component has an effective radius $r_e = 3.71 \pm  0.06$ kpc, a S\'ersic index $n = 0.73 \pm 0.04$, and a position angle of $56^{\rm o} \pm 1^{\rm o}$. There is a broad similarity in the orientation of the optical S\'ersic component and  that  of  the dust disk, PA = $16^{\rm o} \pm 3^{\rm o}$. However, the misalignment is significant, undermining the interpretation that they originate from a common physical structure. Subtracting only the dominant S\'ersic component reveals multiple stellar-mass substructures (Figure \ref{fig:compare_HST_ALMA}); the dust emission is not co-spatial with any of them. The offset from the dust emission peak to the largest stellar mass concentration is $\sim 0\farcs25$. The offset between the centroid of the best-fit S\'ersic component and dust emission is $\lesssim$ 1/6$^{\rm th}$  of the effective diameter of the S\'ersic component. That is, the cold dust emission appears to originate within the geometric centroid of the stellar distribution that could be characterized as the core area of UDF2. However, the region of dust emission is devoid of stellar-mass substructures,  which could be an effect of strong extinction associated with the dust concentration.

None of the remaining four stellar mass substructures coincide with the dust emission; they have a range of offsets of 0\farcs3 to 1\farcs3 ($2-10$ kpc) from its center. These substructures have  corresponding  counterparts  in the {\it HST} $i_{775}$ and $V_{606}$ images that probe the rest-frame emission at $\simeq 1300-2200$ \AA, suggesting that they harbor unobscured star formation. The SFR of these substructures is $< 1$ \Msunyr, less than 0.5\% of the total SFR.

\subsubsection{UDF7}

The dust emission of UDF7 is well-modeled with an elliptical Gaussian and a co-spatial point  source, with the point source a factor of four fainter than the Gaussian component. The Gaussian FWHM is  $0.90 \pm 0.10 \times 0.29 \pm  0.07$ kpc in size, containing $0.58 \pm 0.05$ mJy of integrated flux, whereas the point source is $0.13 \pm 0.03$ mJy (both corrected for the primary beam attenuation). This is an example where the $uv$-based analysis is necessary to measure the size of the dust emitting region, as the minor axis of the Gaussian is comparable in size to  the native beam.  After subtracting the two components, no residual peak $\gtrsim$ 3$\sigma$ remains (local $\sigma$ = 17.6 \uJyperbm). 

The $H_{160}$ image shows complex morphology, requiring five components to model using GALFIT (Appendix A). The dominant S\'ersic component has $r_e = 3.95 \pm 0.04$ kpc and $n = 1.17 \pm 0.01$, with a $\simeq$ 0\farcs1 offset from the centroid  of the dust emission. Nevertheless, given that the dust emission is very compact, it originates entirely from the central  area  of the stellar  mass  concentration.  There are five distinct rest-UV-bright clumps identifiable in the $i_{775}$ and $V_{606}$ images, with counterparts in every band out to $H_{160}$ indicating considerable underlying stellar mass. No cold dust feature is observed at the positions of these star-forming clumps. The dominant S\'ersic component is marginally detected in the $V_{606}$ image, but it is fainter than the UV star-forming clumps. Again, the unobscured star formation in UDF7 contributes $< 1$ \Msunyr\ to the global star formation. In the cases of UDF2 and UDF7, the presence of compact dust emission surrounded by multiple rest-frame UV components is notably similar to those reported by, e.g., \citet{GomezGuijarro18} in six intensely star-forming galaxies at $z \sim 4.5$.

\begin{figure*}[ht]
\figurenum{2}
\centerline{\includegraphics[width=\textwidth]{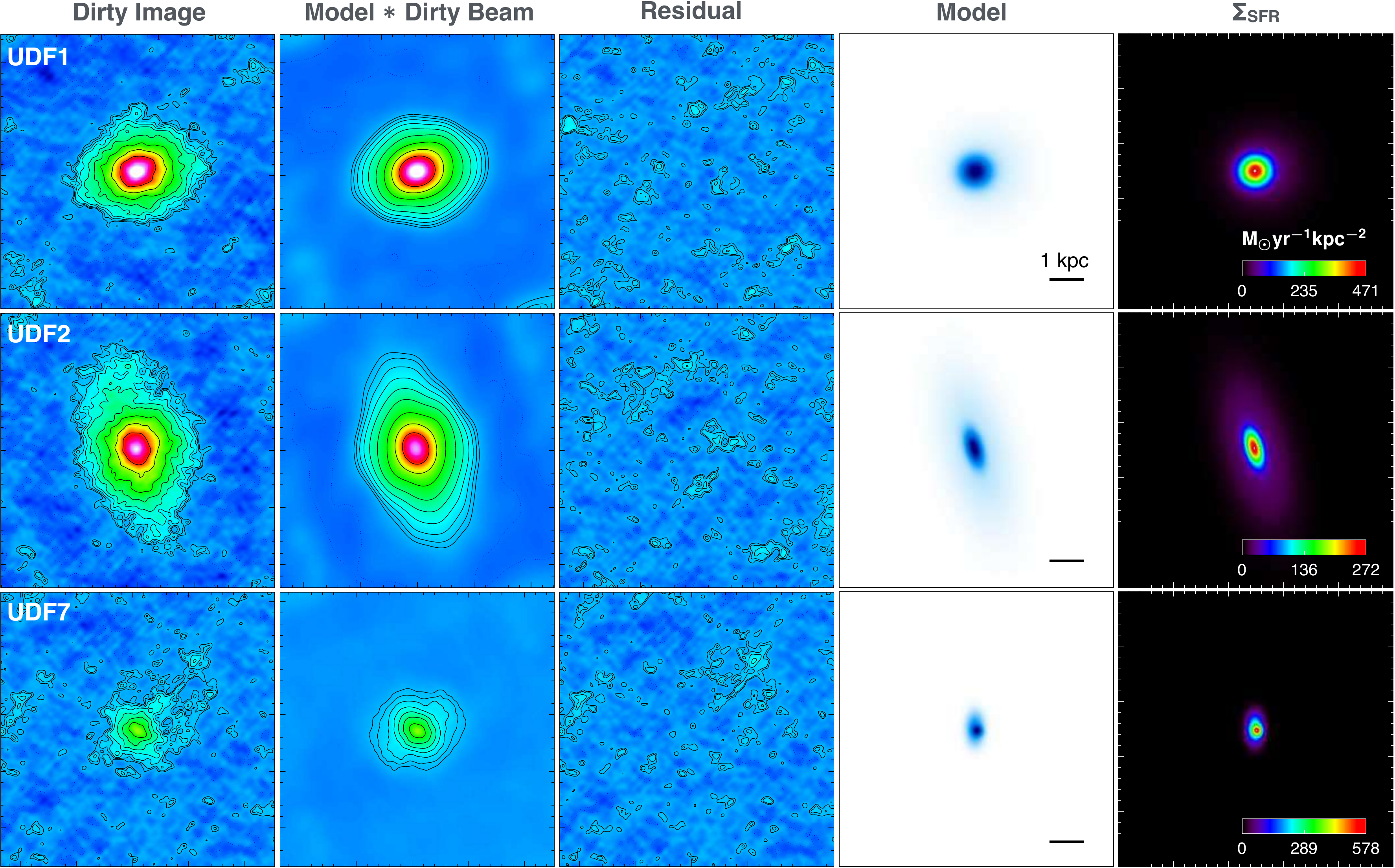}}
\caption{The {\it uv}-plane modeling of HUDF galaxies. From left to right: the dirty image; source model convolved with the dirty beam; model-subtracted dirty image; source model; and the \SFRSD\ constructed from the model. Each image cutout is $1'' \times 1''$; north is up, east is on the left; the contours are [$-1.5, 1.5, 1.5^{1.5}, 1.5^{2}, ...]\times \sigma$; negative contours are dashed; the ALMA beam, not shown is $42 \times 30$ mas, i.e., $\simeq10^3$ beam area in each cutout. All sources are well-described by the models; the resulting residual noise maps are uniform, indicating the lack of substructures. These images are sensitive to SFRs of $\simeq1-3$ \Msunyr\ per 200-pc beam ($1\sigma$). Substructures at this scale can contain no more than $\simeq1, 1,$ and 7\% of the total SFR in UDF1, UDF2, and UDF7 ($3\sigma$), respectively. \label{fig:GILDAS_HUDF}}

\end{figure*}
\begin{deluxetable*}{lccccc}
\tablecaption{UDF source decomposition in the $uv$ plane \label{tab:Comp_UDF}}
\tablehead{
\colhead{ID} & \colhead{Component}  & \colhead{RA} & \colhead{Dec} & \colhead{Flux} & \colhead{Size}\\
\colhead{} & \colhead{} & \colhead{($''$)} & \colhead{($''$)} & \colhead{($\mu$Jy)} & \colhead{(mas)}
}
\startdata
UDF1 & Total  &    03:32:44.033  & 	$-$27:46:35.960				& $ 3407 \pm  226 $ &  \\
	  & 1	      & $   \phantom{-}0.003 \pm 0.001 $  & $  \phantom{-}0.000 \pm 0.001 $  & $ 1844 \pm   159 $  & $   86 \pm   4 \times   82 \pm   4 $ \\
	  & 2	      & $   -0.027 \pm 0.009 $  & $  \phantom{-}0.005 \pm 0.007 $  & $  1562 \pm   159 $  & $  247 \pm  26 \times  228 \pm  24 $ \\
UDF2 & Total  &    03:32:43.529 	   & 	$-$27:46:39.275	& $ 2797 \pm 94\phantom{0} $ \\
 	  & 1	 & $  \phantom{-}0.000 \pm 0.004 $  & $  \phantom{-}0.000 \pm 0.007 $  & $ 1999 \pm   81\phantom{0} $  & $  457 \pm  21 \times  190 \pm  10 $ \\
	  & 2	 & $   \phantom{-}0.007 \pm 0.001 $  & $  \phantom{-}0.007 \pm 0.002 $  & $  798 \pm   48 $  & $   99 \pm   5 \times   44 \pm   4 $ \\
UDF7 & Total  &    03:32:43.326   & 	$-$27:46:46.963				& $  704 \pm   56 $ \\
 	  & 1  & $  \phantom{-}0.004 \pm 0.004 $  & $ \phantom{-}0.001 \pm 0.004 $  & $  576 \pm   48 $  & $  112 \pm  13 \times   36 \pm   9\phantom{00} $ \\
 	  & 2  & $  -0.005 \pm 0.003 $  & $ -0.006 \pm 0.002 $  & $   128 \pm   28 $   & $  ... $
\enddata
\tablecomments{Source IDs are from \citet{Dunlop17}. RA and Dec of each component are tabulated as offsets in arcseconds relative to the position (ICRS) in the corresponding `Total' row. Size are FWHM; dots in the size column indicate an unresolved component. Fluxes are 870 \micron\ integrated flux, corrected for the primary beam attenuation.\\}
\end{deluxetable*}

\subsection{Smoothness of the Cold Dust Distributions}\label{sec:results_smoothness}

Our high-fidelity, extinction-independent observations reveal the central region of intense star formation in each galaxy to be morphologically smooth --- in the sense that only the dominant disk-like dust concentration is present with no additional substructure or clumpy appearance. Inspection of Figure \ref{fig:GILDAS_HUDF} indicates that the residual maps of UDF2 and UDF7 only contain one residual peak above 2.75$\sigma$ and no peak above 3.37$\sigma$ (4$^{\rm th}$ and 5$^{\rm th}$ solid contours, respectively); the residual map of UDF1 contains no peak above 2.75$\sigma$. Given that each cutout contains $\approx 10^3$ beams, we expect up to $\simeq3$ stochastic $3\sigma$ peaks from ideal Gaussian noise. The underprevalence of such peaks provides a conservative limit on the presence of star-forming substructures.

We quantify the upper SFR limit of substructures by scaling the local 870 $\mu$m residual noise per beam to the total 870 $\mu$m flux, which corresponds to the spatially-integrated SFR estimated from fitting the SED with the {\it Spitzer}, {\it Herschel}, and ALMA photometry. While this approach assumes that the dust  temperature and IMF do not vary within each source, it implicitly takes into account  the  dust  temperature  variations  among the galaxies. The $1\sigma$ sensitivities to SFR within our 200-pc beam at the positions of UDF1, 2, and 7 are 1.9, 0.9, 2.8 \Msunyr, respectively. That is, the UV-bright clumps can contain no more than $\simeq$ 1, 1, and 7\% respectively (3$\sigma$ upper limits, given the total SFRs in Table \ref{tab:HUDF_sourcetable}) of the total star formation in these systems.

Our SFR sensitivity to clumps that are intrinsically larger than the 200-pc beam (e.g., $\simeq1$ kpc) has to be scaled correspondingly \citep[cf. ][]{Zanella18}, as we are limited by the surface brightness sensitivity of our observations. We note that this is a potential issue of the surface brightness limitation and not due to the maximum recoverable scale of the interferometer, which has been shown to be larger than our galaxies in Section \ref{sec:obs_ALMA}. However, we argue against missing star-forming clumps in this manner with the new findings on the intrinsic clump sizes via gravitational lensing in Section \ref{sec:discuss_lensed}, comparison with the substructures typically found in SMGs in Section \ref{sec:discuss_H19}, and with the new model predictions on intrinsic clump sizes in Section \ref{sec:discuss_model}.

\subsection{Maps of Star Formation Rate Surface Density}\label{sec:results_SFRSD}

Maps of the star formation rate surface density, \SFRSD, are useful for, e.g., combining with $\Sigma_{M*}$ measurements to study the spatially-resolved star-formation efficiency, as well as predicting the IR SED of the galaxy \citep{Rujopakarn13}. We convert the ALMA dust continuum map into one of \SFRSD\  by scaling the 870 $\mu$m flux to the total in the same way as we placed limits on the clumpy star formation above. Specifically, we constructed a noise-free model for each source from the Gaussian components listed in Table \ref{tab:Comp_UDF}. \SFRSD\ in each model pixel is given by $\Sigma_{\rm SFR} = {\rm SFR}_{\rm total}(S_{\rm 870,pixel}/S_{\rm 870,total})/\Omega_{\rm pixel}$, where SFR$_{\rm total}$ is from $24 - 1300$ $\mu$m SED fitting and $\Omega_{\rm pixel}$ is the pixel area in kpc$^2$. As the sources are well described by the Gaussian models, this approach provides a measure of the \SFRSD\ that is free of fictitious boosting due to noise fluctuations that could interfere with the interpretation of the maps. The \SFRSD\ maps of HUDF galaxies are  shown in Figure \ref{fig:GILDAS_HUDF}.  

These maps indicate that the most intense star formation originates from compact regions: areas with \SFRSD\ $\geqslant 100$ \Msunyrkpcsq\ are limited to the central $0.4-1.0$ kpc$^2$ in these galaxies. A majority of the surface area of the dust distribution (i.e., in the sense of an area larger than those enclosed by the FWHM) harbors \SFRSD\ $\gtrsim 1$ \Msunyrkpcsq\ and possibly drives strong outflows \citep[e.g.,][]{Newman12, Bordoloi14}, to be confirmed with spatially-resolved kinematics.

While this indirect method to produce \SFRSD\ maps carries considerable uncertainties, it is the only tracer capable of 200-pc resolution unaffected by dust extinction. If $A_V$ can indeed be as extreme as $\sim100$ mag reported in SMGs \citep{Simpson17}, this precludes any optical  or near-infrared avenues even with the {\it JWST}: the corresponding  $A_{\rm Pa\alpha}$  would  be  $\sim$ 15  mag  and $A_{\rm Br\alpha}$ would still be $\sim$ 2 mag \citep{RiekeLebofsky85}.  An example illustrating the limitations of optical/near-IR tracers is  \citet{Nelson19}'s report of H$\alpha$ emission in a $z \simeq 1.25$ galaxy with a conspicuous void in the middle where, however, the dust continuum indicates intense star formation \citep[see also, e.g., ][]{FS18}. We observe a similar effect in the central region of UDF2, where the $H_{160}$ emission (rest-frame 0.43 $\mu$m) appears suppressed at the location of dust emission. ALMA remains the only high-resolution extinction-independent tracer until the era of next-generation radio facilities.

We note that the implicit assumptions that the dust temperature and IMF do not vary spatially within each galaxy are likely to be challenged with future observations. The variation of the dust temperature is straightforward to constrain with additional spatially-resolved observations across the dust spectrum. The IMF variation is less so. Some recent results hint at a more top-heavy IMF at the sites of intense starbursts locally \citep{Schneider18, Motte18} and at $z \sim 2-3$ \citep{Zhang18}. If confirmed to be a common occurrence in typical SFGs at $z \sim 3$, it is possible that the intense starburst near the centers of these galaxies could have systematically top-heavy IMFs compared to the outskirts, thereby lowering the central \SFRSD.

\section{DISCUSSION}\label{sec:discuss}

\subsection{Spatial distribution of cold dust}\label{sec:discuss_distribution}

While two components were needed to model the cold dust emission of our HUDF galaxies, we did not require them to be co-spatial --- their centroids were free parameters. However, the best fit models, yielding uniform noise residual maps, indicate that they are co-spatial within $\simeq 10$ mas (Table \ref{tab:Comp_UDF}), comparable to the corresponding positional uncertainties. In the case of UDF2, where the source elongation is very pronounced, the position angles of the two Gaussians are in excellent agreement: $14.6^{\circ} \pm 1.7^{\circ}$ and $15.7^{\circ} \pm 2.9^{\circ}$. The co-spatiality, in effect, creates a linear superposition of Gaussians, which is consistent with our expectation that such a superposition can accurately represent a single S\'ersic profile \citep[][discussion in Section \ref{sec:obs_GILDAS}]{HoggLang13}.

On the optical side, we highlight two challenges in interpreting the {\it HST} morphologies before further discussion of the ALMA/{\it HST} comparison. First, while disturbed $H_{160}$ morphologies are observed in all three of our HUDF galaxies (including the visually innocuous UDF1), the origin of such disturbances is inconclusive with the possibilities ranging from in-situ disk instability or patchy dust extinction, to major and minor mergers (or any combinations of these). Second, we cannot ascertain that the rest-frame UV and optical substructures in angular proximity to the galaxies are physically associated with the galaxies (or simply foreground or background sources). Definitive confirmation of associations will require spatially-resolved spectroscopy at resolution comparable to our ALMA observations (tens of mas) from next-generation optical facilities. The following discussion assumes that the rest-frame optical emission modeled with GALFIT (Figure \ref{fig:A1} in Appendix A) and the rest-frame UV emission circled in Figure \ref{fig:compare_HST_ALMA} are associated with our galaxies.

For UDF2 and 7, where UV star-forming clumps are seen, they are spatially distanced by  $\simeq 2-10$ kpc from most of the star formation. The dislocation is a common occurrence in typical $z \sim 2 - 3$ massive star-forming galaxies \citep{Rujopakarn16, Dunlop17} as well as in submillimeter-bright galaxies at $z \sim 3 - 4$ \citep{Hodge16, GomezGuijarro18} and presents a fundamental limitation to the application of an energy-balancing argument \citep[e.g.,][]{daCunha15}  to estimate their total, mostly obscured, SFR. This effect is also seen in the large dispersion of the infrared excess (IRX) for ALMA-selected galaxies, ranging from the level consistent with the \citet{Calzetti00} extinction law to $1-2$ orders of magnitudes above it  \citep{McLure18}, similarly representing an obstacle to the application of the IRX-$\beta$ method \citep[$\beta$ being the rest-frame UV continuum slope, $S_{\lambda} \propto \lambda^{\beta}$;][]{Meurer99} to estimate the total SFR in luminous SFGs typical at this epoch \citep[see also, e.g.,][]{Hodge16, Simpson17, GomezGuijarro18}.

In all three galaxies, we observe offsets of $0\farcs1-0\farcs2$ between the bulk of star formation and the centroids of the stellar mass distributions. While these offsets are comparable  to  the 0\farcs15 rms of the  registration of {\it HST} astrometry to the ICRS in GOODS-S, the median offset between {\it HST} and ICRS is less than 10 mas \citep[][M. Franco et al., in preparation]{Rujopakarn16}. Furthermore, the dust location either coincides with the optical AGN (UDF1) or a dark area that may represent strong extinction (UDF2 and 7). These results suggest that the offsets are real, but, e.g., {\it JWST}/NIRCam imaging will be required for a definitive confirmation. Nevertheless, given the  S\'ersic sizes  and the general proximity of the  dust emission to the S\'ersic centroids, it can be  established with the current imaging that the dust emitting regions are embedded in the S\'ersic disk.

While the dust emitting regions are visually more compact than the $H_{160}$ extent \citep[also reported by, e.g.,][]{Chen15, Barro16, Fujimoto18, Hodge19}, it is possible that the $H_{160}$ images do not reflect the true stellar mass distribution of these galaxies due to strong dust extinction, especially in the cases of UDF2 and 7. Again, {\it JWST} and next-generation facilities will be required for a definitive study of the relationship between \SFRSD\ and $\Sigma_{M_*}$.

Regardless of the  fiducial  stellar  mass  distribution at the location of intense star formation, the ongoing episode of star formation is capable of doubling the {\it global} stellar mass within $0.2 - 0.4$ Gyr, and the newly formed stellar mass will likely lie within the innermost $\lesssim$ 0.5 kpc. Whether these episodes of star formation are capable of forming bulges will depend critically on the nature of any feedbacks, e.g., star-formation driven outflows, which remain to be characterized in a spatially-resolved manner in typical SFGs at $z \sim 3$.

The intensely star-forming region of UDF7 has previously been reported by \citet{Rujopakarn18} to be co-spatial with the location of excess radio emission signifying the location of an AGN, which has been localized to $\lesssim100$ pc using high-fidelity VLA imaging at 6 GHz \citep{Rujopakarn16}. We confirm with the improved ALMA imaging in this work that the AGN is co-spatial with intense star-formation at this scale, consistent with  a picture of in-situ bulge formation with co-spatial and contemporaneous growth of supermassive black holes. Similarly, the bright point source required to model the $H_{160}$ morphology of UDF1, which likely represents the AGN emission, is also co-spatial with the dust emission.

Lastly, the smoothness of the cold dust emission is remarkable considering the complex and disturbed the rest-frame UV emission, which indicates that a galaxy-wide disturbance has occurred recently. This highlights how quickly the cold gas is capable of resettling into the star-forming disk following such a disruptive event. While the emergence of a kinematically-ordered disk is anticipated from simulations of, e.g., gas-rich mergers typical in the early Universe after a $\sim$Gyr \citep{Robertson06, Hopkins09} and rotationally-supported disks are observed in local merger remnants \citep{Ueda14}, the dust disk resettlement timescale of $\lesssim0.1$ Gyr implied by the disturbed UV-bright morphology (i.e., the lifetime of OB stars) is not expected from the aforementioned models and suggests that disruptive event in gas-rich environment may be far more dissipative than previously thought. Without drawing a conclusion that our HUDF galaxies are mergers, we nevertheless point out that recent merger simulations such as those of \citet{Fensch17} that take into account the environment and dynamics typical of the high-$z$ ISM  have found the gas resettling time to be $\sim0.1-0.2$ Gyr. This highlights the importance of characterizing the origin of the disruptive events in these galaxies.

\subsection{Clumplessness vs. previously reported clumps}\label{sec:discuss_litcompare}

We now put the observed clumplessness of the HUDF galaxies in context by comparing with the previously reported star-forming clumps at various wavelengths, and with the clumps identified from high-resolution images obtained via gravitational lensing. The star formation substructures found in the high-fidelity ALMA observations of SMGs by \citet{Hodge19} deserve special attention and will be discussed separately in Sections \ref{sec:discuss_H19}.

\subsubsection{Previously reported clumps at various wavelengths}\label{sec:discuss_field}

We will first focus on the recent reports of star-forming substructures at $z \simeq 1-3$ from optical and near-IR observations. As clump properties depend strongly on the SFR of their host galaxy, we will discuss the results in the following descending order of SFR: \citet{Genzel11}, \citet{Wisnioski12}, \citet{Swinbank12}, \citet{Guo18}, and \citet{Soto17}.

The \citet{Genzel11} sample of five typical SFGs is the most similar to our HUDF galaxies in terms of redshift, SFR, and stellar mass:  $z \sim 2.3$;  SFR $\simeq 70 -  180$ \Msunyr;  log($M_*$/\Msun) $\simeq 10.3 - 11.0$. The galaxies are drawn  from  the  spatially-resolved  spectroscopic sample of H$\alpha$ emission using VLT/SINFONI \citep{FS09}; additional AO-assisted SINFONI observations were conducted to achieve a typical resolution of $0\farcs2$. \citep{Genzel11}. More than 20 kpc-sized star-forming clumps are found in five galaxies with individual clumps the sites of SFRs of $3 - 40$ \Msunyr\ (typically $10-20$ \Msunyr), which are $6-20$\% of the total SFR in each galaxy (we will refer to this percentage as the `fractional SFR' hereafter) based on their H$\alpha$ emission. Such star-forming clumps, especially considering their fractional SFR, would have been very strongly detected if any existed in our HUDF galaxies. 

\citet{Wisnioski12} describe a sample at a slightly more recent cosmic epoch, $z \simeq 1.3$, and less strong SFRs, $20 - 50$ \Msunyr, but similar stellar masses, log($M_*$/\Msun) $\simeq 10.7 - 11.0$, i.e., those assembling their stellar mass at relatively later time compared to the \citet{Genzel11} sample. They find eight clumps in three galaxies from spatially-resolved H$\alpha$ observations using Keck/OSIRIS. Overall, their clumps contain a median SFR of 4 \Msunyr, which is 13\% in terms of the median fractional SFR. The \citet{Swinbank12} sample of AO-assisted SINFONI spatially resolved H$\alpha$ spectroscopy of nine galaxies drawn from a narrow-band H$\alpha$ imaging survey is at a similar redshift of $z \approx 0.8$ and 1.4 but with a much lower median stellar mass of log(M$_*$/M) $\approx$ 10.1. With the total host SFRs of $1 - 10$ \Msunyr, the SFR of individual  clumps is also smaller at $0.5 - 2.9$ \Msunyr, implying an even larger fractional SFR of $\sim$ 25\%. Our imaging sensitivity is not sufficient to detect the level of SFRs in the individual \citet{Wisnioski12} and \citet{Swinbank12} clumps if they are at $z = 3$ (possibly detecting $2 - 3$ clumps from the former). However, if the fractional SFRs in our targets were as large as in these hosts, they would also have been well detected.

For completeness, we also consider the comprehensive selections of rest-frame UV clumps by \citet{Soto17} and \citet{Guo18} using the {\it HST} rest-frame far-UV and near-UV imaging, respectively. At $z = 1.5 - 3.0$, the \citet{Guo18} catalog contains 1083 clumps in 501 galaxies; the selection bands are the CANDELS imaging at $V_{606}$ for $1.0 \leqslant z \leqslant 2.0$ and $i_{775}$ for $2.0 \leqslant z \leqslant 3.0$ galaxies. These clumps have a median log($M_*$/\Msun) of 8.4 (cf. stellar mass of gravitationally lensed star-forming clumps in Section \ref{sec:discuss_lensed}) and a median SFR of the clumps and hosts of 1.2 and 22.2 \Msunyr, respectively (7\% fractional SFR). At lower redshifts ($z = 0.5 - 1.5$), the \citet{Soto17} sample contains typical SFGs selected from the {\it HST} F225W, F275W, and F336W imaging, with median stellar mass and SFR of $\simeq$ 10$^7$ \Msun\ and 0.29 \Msunyr, respectively; again, with a fractional SFR of 5\%. We would not be able to detect these unobscured UV clumps if they are at $z = 3$, as evident from the optical comparisons in Section \ref{sec:results_HST_compare}. Two UV clumps in UDF7, the brightest in the optical among our three galaxies, are actually cataloged by \citet{Guo18}, but no dust emission is observed at their locations.

The sensitivity and resolution of sub/millimeter observations have only started to approach optical observations with the advent of ALMA, and still can only reach the more luminous SFR regimes. A majority of high-resolution (e.g., resolution $\lesssim$ 0\farcs2)  studies  of  unlensed  galaxies with ALMA were carried out on SMGs selected from single-dish surveys  \citep[e.g., the JCMT and APEX surveys of COSMOS, ECDFS, and UDS at 850 \micron\ and 1.1 mm by][]{Scott08, Weiss09, Geach17}. These surveys pushed their sensitivities to the limit of  the confusion noise,  typically $\sim$ 1 mJy,  yielding reliable detections above sub/millimeter fluxes of $\sim$ 4 mJy, and hence the parent samples for the following observations are populated by luminous SFGs.

\citet{Iono16} carried out 350 GHz ALMA observations of three sources from the AzTEC survey in COSMOS (AzTEC 1, 4, and 8; $z  = 3.12 - 4.34$) with SFR $\sim 1300-2800$ \Msunyr\ at angular resolutions of $15 - 50$ mas. They reported spatially resolving the sources into multiple substructures; two of the brightest ones in AzTEC-1 (each with SFR $\simeq 50$ \Msunyr) were  confirmed with CO($4-3$)  kinematics by \citet{Tadaki18}. \citet{Hodge16} observed 16 sources from the ALESS survey \citep{Hodge13} at 354 GHz with 0\farcs16 resolution, revealing them to be extended dust disks with typical diameters of $3.6 \pm 0.4$ kpc. While a majority of the \citet{Hodge16} galaxies appear smooth at their resolution and sensitivity, some do show signs of substructures. \citet{Hodge16} demonstrated with simulations that at moderate S/N of $5 - 10$, smooth disks can be broken up into visually-convincing spurious clumps. However, with improved ALMA data, \citet{Hodge19} have shown many of the clumps in these galaxies to be real (Section 4.2.3 is dedicated to discussing this result). Similarly, dust continuum and [CII] imaging at 30 mas resolution by \citet{Gullberg18} of four $z \simeq 4.4 - 4.8$ SMGs visually show multitudes of substructures. As with Hodge et al., they caution that there is a significant probability that this result may be consistent with a smooth disk. These studies are reminders that high fidelity observations are required in pushing the resolution and sensitivity of interferometric imaging to search for star-forming substructures.

As a result of the confusion limit, even the comparably faint single-dish-selected SMGs are brighter in the far-IR than typical SFGs: they have typical S$_{870 \mu m}$ of $5-15$ mJy \citep{Weiss09, Hodge13}, whereas massive main-sequence SFGs have S$_{870 \mu m}$ of about $1-3$ mJy at $z \sim 3$ \citep[][and those in this work]{Scoville17}. As discussed in the following section, observations of gravitationally-lensed main-sequence SFGs at $z \sim 2-3$ also provide no evidence of substructures. To date, there is no confirmed report of substructures in the bulk of star formation in main-sequence SFGs at $z \sim 2-3$. 

\subsubsection{Clumps in gravitationally lensed galaxies}\label{sec:discuss_lensed}

Strong gravitational lensing by a galaxy or galaxy cluster preserves surface brightness while spatially magnifying background sources by a factor of as much as a few hundreds. With proper lens modeling, this technique allows current-generation observing facilities to capture high-resolution details in distant galaxies not otherwise feasible. Highly magnifying lenses are often selected from optical surveys of galaxy clusters by identifying bright, extended arcs \citep[e.g.,][]{GladdersYee05, Bayliss12}.

Gravitational lensing provides more tests for clumps at lower redshifts ($z \lesssim 2$) and/or at lower SFRs. Recent observations of these lensed `giant arcs'€™ often achieve physical resolution in the source plane of $\simeq 50-100$ pc. For example, \citet{Sharon12} and \citet{Wuyts14} observed a giant lensed arc RCSGA0327 with {\it HST} and Keck/OSIRIS, a $z = 1.70$ galaxy forming stars at 29 $\pm$ 8 \Msunyr. Source reconstruction indicated at  least seven star-forming clumps, each $300-600$ pc in diameter. In SDSS J1110+6459, a $z = 2.48$ galaxy with a global SFR of 8.5 \Msunyr, \citet{Johnson17a, Johnson17b} reported 27 clumps, each $60 - 100$ pc in diameter and forming stars at a few 10$^{-3}$ \Msunyr\ \citep[see also,][]{Jones10, MenendezDelmestre13, Livermore15, Girard18}.

\citet{Cava18} reported 55 star-forming clumps in the `€˜Cosmic Snake'€™, a galaxy at $z = 1.04$ forming stars at 30 \Msunyr. The background galaxy is lensed into two images: a highly magnified one with magnifications of tens to more than 200 along the namesake arc, and a much less magnified counter-image that has a magnification of 4.5. The physical resolution that  can be reconstructed from the Snake is as small as 30 pc, whereas the resolution from the counterimage is limited to 300 pc (comparable to {\it HST}'s). Notably, the clumps identified from the 30-pc image have an average mass of log($M_*$/\Msun) = 8.0, compared with log($M_*$/\Msun) = 8.7 from the 300-pc reconstruction; the sizes of the clumps identified in the 300-pc image are also twice as large (after correcting for the lensing effect). The differences highlight the fictitious increase in clump size and mass when measured at lower resolution due to clusters of smaller clumps blending together into the appearance of larger clumps, even with other systematics mitigated by  observing the same background source via  a natural lens.

\begin{figure*}
\figurenum{3}
\centerline{\includegraphics[width=\textwidth]{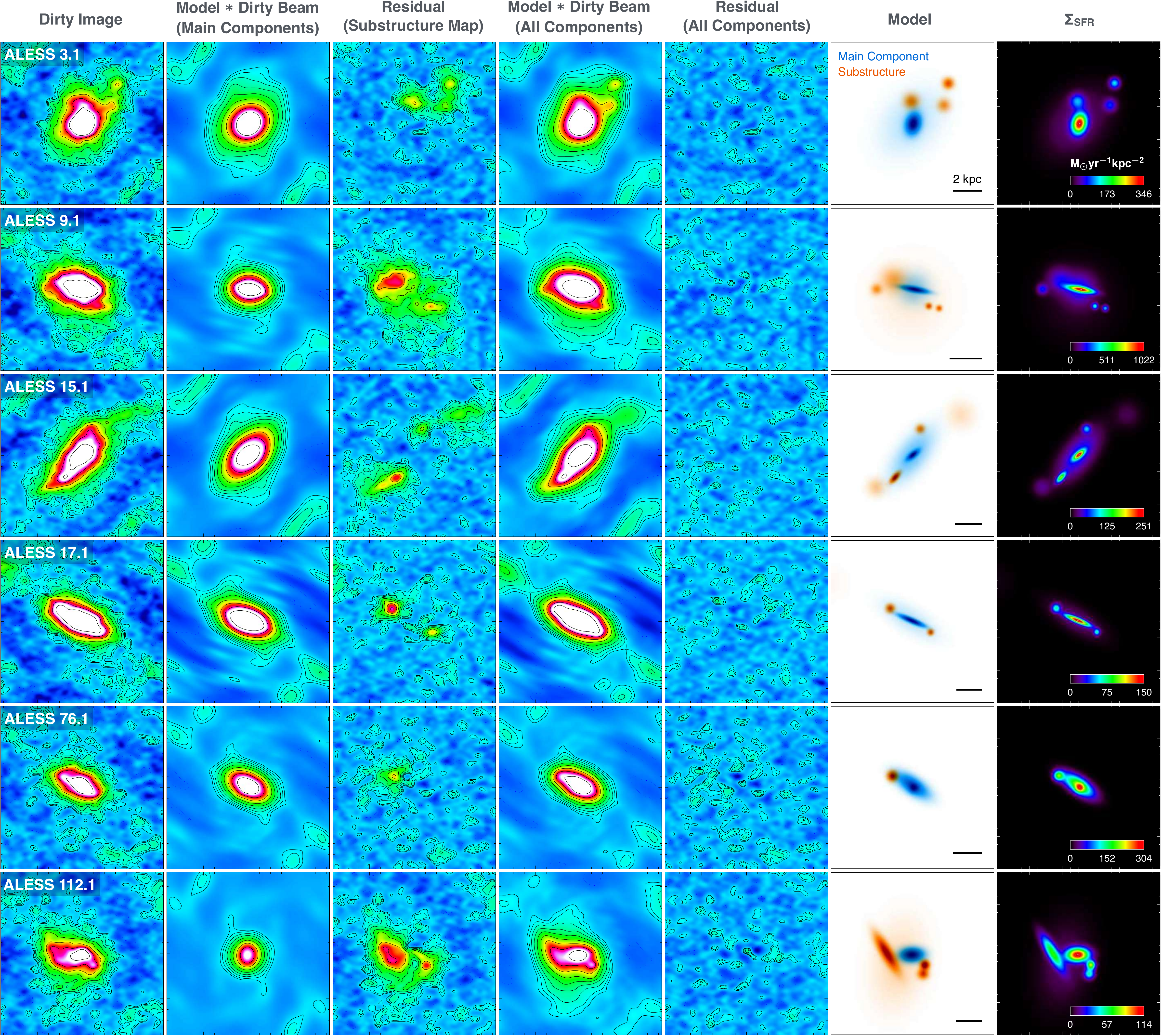}}
\caption{Applying the same analysis method as done with the HUDF galaxies (Figure \ref{fig:GILDAS_HUDF}), we confirm the \citet{Hodge19} findings that  submillimeter-selected galaxies (SMGs) in their samples harbor multiple substructures (Section \ref{sec:discuss_H19}). From left to right: dirty image; model convolved with the dirty beam for main components (those marked as ``main'' components in Table B1 in Appendix B); residual after subtracting the main components, i.e., dirty map of substructures; source model convolved with dirty beam for the all the components; residual after subtracting all components;  source model showing the main components in blue and substructures in red; \SFRSD\ map. Each image cutout is $1\farcs5 \times 1\farcs5''$; north is up, east is on the left; the contours are [$-1.5, 1.5, 1.5^{1.5}, 1.5^{2}, ...]\times \sigma$. Many of the substructures (red components in the model column) in the \citet{Hodge19} SMGs are as bright at $S_{\rm 870}$ as our individual HUDF galaxies.\\
\label{fig:GILDAS_ALESS}}
\end{figure*}

\begin{deluxetable*}{lcccccccccc}
\tablecaption{Submillimeter-selected Galaxies in the \citet{Hodge19} ALESS Sample\label{tab:ALESS_sourcetable}}
\tablehead{
\colhead{ID} & \colhead{RA} & \colhead{Dec} & \colhead{$z$} & \colhead{M$_{\rm *}$} & \colhead{M$_{\rm *, H19}$} & \colhead{$L_{\rm IR}$} & \colhead{SFR} & \colhead{$M_{\rm dust}$} & \colhead{M$_{\rm gas}$} & \colhead{$f_{\rm gas}$}  \\
\colhead{} & \colhead{(deg)} & \colhead{(deg)} & \colhead{} & \colhead{(log\Msun)} & \colhead{(log\Msun)} & \colhead{(log$L_{\odot}$)} & \colhead{(log\Msun)} & \colhead{(log\Msun)} & \colhead{(log\Msun)} & \colhead{} 
}
\startdata
ALESS 3.1   & 53.33964 & $-27.9224$ &  3.374 & $10.2 \pm 0.2$ & $11.30^{+0.19}_{-0.24}$ &  $12.89 \pm 0.03$  &   $ 880 \pm   30$  &  $9.45 \pm 0.04$ &  $10.9 \pm 0.2$ & $0.8 \pm 0.1$ \\
ALESS 9.1   & 53.04722 & $-27.8700$ &  4.867 & $10.8 \pm 0.3$ & $...$ &  $13.33 \pm 0.06$  &   $2430 \pm  160$  &  $9.21 \pm 0.02$ &  $10.7 \pm 0.2$ & $0.4 \pm 0.2$ \\
ALESS 15.1  & 53.38905 & $-27.9916$ &  2.67  & $10.6 \pm 0.2$ & $11.76^{+0.21}_{-0.26}$ &  $12.62 \pm 0.02$  &   $ 470 \pm   10$  &  $9.65 \pm 0.03$ &  $11.1 \pm 0.2$ & $0.8 \pm 0.1$ \\
ALESS 17.1  & 53.03035 & $-27.8558$ &  1.539 & $10.8 \pm 0.2$ & $11.01^{+0.08}_{-0.07}$ &  $12.26 \pm 0.05$  &   $ 210 \pm   10$  &  $9.95 \pm 0.03$ &  $11.4 \pm 0.2$ & $0.8 \pm 0.1$ \\
ALESS 76.1  & 53.38479 & $-27.9988$ &  3.389 & $10.5 \pm 0.3$ & $11.08^{+0.29}_{-0.34}$ &  $12.23 \pm 0.40$  &   $ 190 \pm   90$  &  $9.63 \pm 0.24$ &  $11.1 \pm 0.4$ & $0.8 \pm 0.2$ \\
ALESS 112.1 & 53.20357 & $-27.5203$ &  2.315 & $11.2 \pm 0.2$ & $11.36^{+0.09}_{-0.12}$ &  $12.57 \pm 0.04$  &   $ 420 \pm   20$  &  $9.57 \pm 0.04$ &  $11.0 \pm 0.2$ & $0.4 \pm 0.1$
\enddata
\tablecomments{Source IDs are from \citet{Hodge19}. $M_*$ from our estimate using conventional SED fittings and those from H19 based on the energy-balancing {\tt MAGPHYS} tabulated (except ALESS 9.1 that is affected by blending with a nearby object), details in Section \ref{sec:discuss_SFG_SMG}. We assume a starburst $M_{\rm gas}$/$M_{\rm dust}$ of 30 here. $f_{\rm gas} = M_{\rm gas}/(M_{\rm gas} + M_*)$, adopting our $M_*$ estimates.\\
\phantom{This invisible text is placed here to force a separation between this table and the text/figure below.}}
\end{deluxetable*}

Similar results have been reported by others. \citet{DessaugesZavadsky17} compared luminosities and stellar masses of star-forming clumps identified from unlensed and lensed galaxies from the literature and found a systematic difference. Clump stellar masses from lensed galaxies have  median log($M_*$/\Msun) $\simeq$  7.0 while the median of those from field galaxies is log($M_*$/\Msun) $\simeq 8.9$. This is also reflected in their median luminosity, M$_V$, with distributions peaking at M$_V \simeq -19$ and  $\simeq -17$   for the field and lensed galaxies, respectively. Field identification of clumps primarily relies on {\it HST} imaging with spatial resolution of 0\farcs1, corresponding to   $\simeq 1$ kpc at $z \sim 1-3$.  \citet{DessaugesZavadsky17} suggest that the limited spatial resolution and the propensity of the intrinsically  $\sim 100$ pc star-forming clumps with log($M_*$/\Msun) $\sim 7-8$ to cluster may be responsible for the appearance of giant kpc clumps with log($M_*$/\Msun) $\simeq$ 9. Similarly, \citet{Rigby17} smoothed the model image of the aforementioned SDSS J1110+6459 to simulate the resolution of {\it HST} and found that the intrinsic 27 clumps were blended together into a single disk with S\'ersic  index $n = 1.0 \pm 0.4$  and $r_e = 2.7 \pm 0.3$ \citep[see also a similar result from smoothing low-$z$ galaxy images to simulate $z \sim 2$ galaxies by][]{Fisher17}. It remains an open question whether giant, massive kpc-scale star-forming clumps really do exist and whether clump-clump mergers are capable of producing such massive clumps in their course of evolution. We will revisit the topic of the intrinsic clump sizes from the theoretical perspective in Section \ref{sec:discuss_model}.

While the selection of optically-bright giant arcs in galaxy clusters affords extraordinary magnification to characterize the intrinsic clumps, it also selects hosts with relatively low SFR that are, by necessity, unobscured. The clumps identified are therefore more similar to  rest-frame UV clumps such as those described by \citet{Soto17} and \citet{Guo18}. Again, highly magnified galaxies in the far-IR are needed to probe the substructures in the bulk of star formation in massive main-sequence SFGs such as those of our HUDF targets.

A notable far-IR search for clumps in main-sequence SFGs at $z > 2$ was conducted on the strongly lensed ``Eyelash''. \citet{Swinbank10} reported four clumps of $100 - 300$ pc diameter in this gravitationally-lensed typical SFG with a total unlensed SFR of $210 \pm 50$ \Msunyr\ at $z = 2.33$. The reported star-forming substructures are clearly visible at $4-7\sigma$ in their 870 $\mu$m image from the Submillimeter Array \citep[SMA; ][]{Swinbank10}. However, recent ALMA observations at multiple bands with higher sensitivity than those from the SMA indicate that the Eyelash is morphologically smooth \citep[][Ivison et al., in preparation]{Falgarone17}. While this is a gravitationally lensed system, the disagreement is unrelated to the lens, but rather to the challenges of  interferometric imaging. 

The most prolific approach to identify far-IR lenses is to conduct large area shallow surveys for far-IR-bright sources. This is because the bright end of the far-IR number counts plummets rapidly above, e.g., S$_{500 \mu m}$ of 100 mJy, such that the fraction of lensed galaxies above this range is near unity \citep{Negrello17}. The largest samples of far-IR lenses were identified from surveys such as the South Pole Telescope (SPT) 87 deg$^2$ surveys at 1.4 and 2.0 mm \citep{Vieira10} and the {\it Herschel}-ATLAS 570 deg$^2$ survey at $100-500$ $\mu$m \citep{Eales10}. Far-IR lenses selected from this approach tend to be luminous SFGs at higher redshifts due to the combination of the aforementioned efficient lens selection at bright far-IR flux and the larger effect of the negative K-correction at  $z \gtrsim 2$.  For example, the \citet{Spilker16} sample of SPT lenses has a median redshift $z = 4.3$ and median SFR of 1100 \Msunyr.

The highest resolution and fidelity observations of a lensed system to date are those of SDP.81 at $z = 3.0$, identified from {\it Herschel}-ATLAS, as a part of the ALMA long baseline campaign \citep{ALMAPartnership15}, which revealed multiple substructures. Given a Toomre parameter Q of $\simeq$ 0.3, SDP.81 is likely a merger system that drives intense SFR of $\simeq$ 500 \Msunyr\ at high star formation efficiency \citep{Swinbank15, Dye15}. \citet{Swinbank15} and \citet{Tamura15} reported 5 and 35 star-forming clumps, respectively, from the same data; this apparent disagreement illustrates the challenge in clump identification even with high fidelity data. As with  interferometric imaging of star-forming substructures in field galaxies (Section \ref{sec:discuss_field}), low S/N clumps in lensed sources are subject to the same pitfalls of a smooth disk being broken into clumps, because gravitational lensing preserves surface brightness. The $\geqslant 5\sigma$ clumps reported by \citet{Swinbank15} are $200-300$ pc in  diameter after correcting for lensing effects. Such clumps should be well detected in our ALMA data, given their large fractional SFR \citep[Figure 1 of][]{Swinbank15} and the comparable sensitivity  between  our  observations  and  those  of  SDP.81 (10 vs. 11 \uJyperbm\ rms at Band 7, respectively) to the rest-frame $240-250$ $\mu$m dust continuum. However, SDP.81 is $\simeq 5$ mJy at 880 $\mu$m after correcting for the lensing magnification \citep[][continuum magnification assumed]{Dye15}, i.e., more similar in the far-IR flux to SMGs (Section \ref{sec:discuss_H19}) and $2-3$ times brighter than our HUDF galaxies. While challenging, future high-fidelity observations similar to those of SDP.81, but on lensed sources identified to be representative of main-sequence SFGs, are needed to study intrinsic clumps in the typical SFG population.

\begin{figure}
\figurenum{4}
\centerline{\includegraphics[width=\columnwidth]{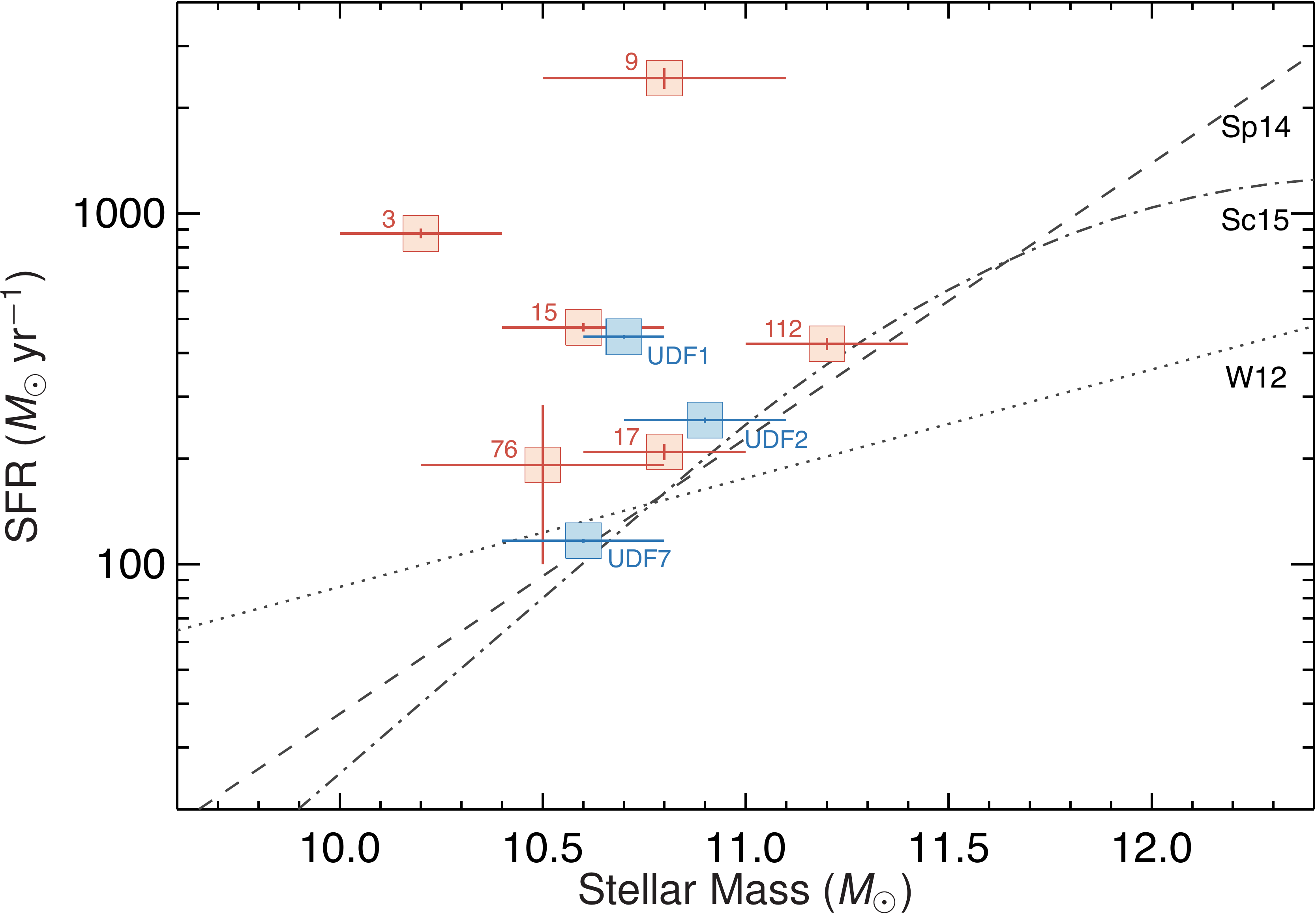}}
\caption{Typical star-forming galaxies (SFGs) in the HUDF (blue squares) lie securely within the scatter of the main sequence of SFGs, so do most of the submillimeter-selected galaxies (SMGs) from ALESS in the \citet{Hodge19} sample (red squares), even considering a very conservative range of possible stellar masses. Both samples also form stars at comparable rates. The main reason that ALESS SMGs are detected in the single-dish survey (which also covers the HUDF) appears to be the larger dust mass. The dotted, dashed, and dot-dashed lines are the main sequence parameterizations at $z = 3.0$ by \citet{Whitaker12},  \citet{Speagle14}, and \citet{Schreiber15}, labeled W12, Sp14, and Sc15, respectively.
\label{fig:mainseq}}
\end{figure}

\subsubsection{Clumplessness in typical SFGs vs. clumps in SMGs}\label{sec:discuss_H19}

\citet{Hodge19}, hereafter H19, found multiple clump-like star-forming substructures in six SMGs based on high-fidelity ALMA observations (we use the terms `clump' and `substructure' interchangeably in the following discussion). Observationally, the difference between their and our samples is primarily the far-IR flux densities due to the selection from single-dish vs. ALMA deep field observations. As all six H19 SMGs are clumpy, whereas all three SFGs in our sample are clumpless, H19 provides a direct comparison sample to investigate the difference between the two.

The six SMGs are at $z = 1.53 - 4.86$ in the ECDFS. As with ours, their observations were at 870 $\mu$m. They were first identified from a single-dish survey \citep[LESS,][]{Weiss09} and followed up  with  ALMA  during  Cycle  0  at $1''- 3''$ resolution  \citep[ALESS,][]{Hodge13}, from which 16 SMGs were selected for 0\farcs16-resolution observations by \citet{Hodge16}. From these 16 SMGs, six among the brightest (to minimize the observing time) were selected for 70 mas observations with rms sensitivities of $22 - 26$ \uJyperbm, i.e., same band, approximately a factor of two larger synthesized beam diameter and a factor of two less sensitive than our work.

For a rigorous comparison, we took the pipeline product of the H19 archival data (ALMA\#2016.1.00048.S) and analyzed the calibrated visibilities in the $uv$ plane with GILDAS using the identical procedure as for our data (Section \ref{sec:obs_GILDAS}). We fitted all components of the SMGs simultaneously with all parameters of every component being free. Components were added to the model one-by-one in an increasing order of free parameters (e.g., point, circular Gaussian, elliptical Gaussian); the simultaneous fit was re-run each time as necessary to achieve a uniform residual after subtraction, shown in Figure \ref{fig:GILDAS_ALESS}. These components and their parameters are listed in Table B1 in Appendix B. The differences between the two data sets are evident: while the HUDF galaxies are well modeled by two cospatial Gaussians, the H19 SMGs require multiple additional small components, confirming the presence of dust substructures reported by H19.

After modeling all component of each source in the $uv$-plane, we identify their far-IR substructures by subtracting the dominant components, defined by those positions coinciding with the peak emission. As shown in Table B1, there is little ambiguity as to which components are the dominant ones in all cases except that of ALESS 112.1, where there are two dominant components with integrated fluxes of $2.0 \pm 0.1$ and $2.3 \pm 0.2$ mJy; we subtract the former because it is at the location of the peak flux of the source. The substructure maps are shown in Figure 3 along with their source models and the \SFRSD\ maps. Our images are similar to those in H19 from modeling in the image plane; the general agreement shows that the differences between the HUDF and H19 samples are not an artifact of the reductions.

Assuming a constant dust temperature and IMF, i.e., that the SFR varies linearly with the rest-frame $150-350$ $\mu$m dust emission, the typical fractional SFRs in substructures in the H19 SMGs range from 4 to 10\%, with two substructures being as large as $30-36$\%. Clearly, this is higher than the $1-7$\% ($3\sigma$) upper limits in our HUDF galaxies. The average integrated flux of the H19 galaxies from our measurement is 8.0 mJy, whereas the average for HUDF galaxies is 2.3 mJy (cf. Tables \ref{tab:Comp_UDF} and B1). The presence of dust substructures and brighter far-IR fluxes are the two concrete observational differences between the HUDF and H19 samples.

While our analysis agreed with H19 in the presence of substructures in H19 SMGs, there are some differences in identity, size, and flux of the substructures between our morphological decomposition and that of H19. The fraction of total flux contained in individual substructures reported by H19 is smaller ($2-8$\%) than we have found ($2-36$\%). These differences could be attributable to, e.g., the choice of model for the main component (single exponential vs. nested Gaussians) or the differences in the residual noise characteristics between our substructure maps in Figure 3 and those of H19 (their Figure 3, rightmost column), which serves to highlight the range of possible answers that can arise from the methodological differences.

\begin{figure}
\figurenum{5}
\centerline{\includegraphics[width=\columnwidth]{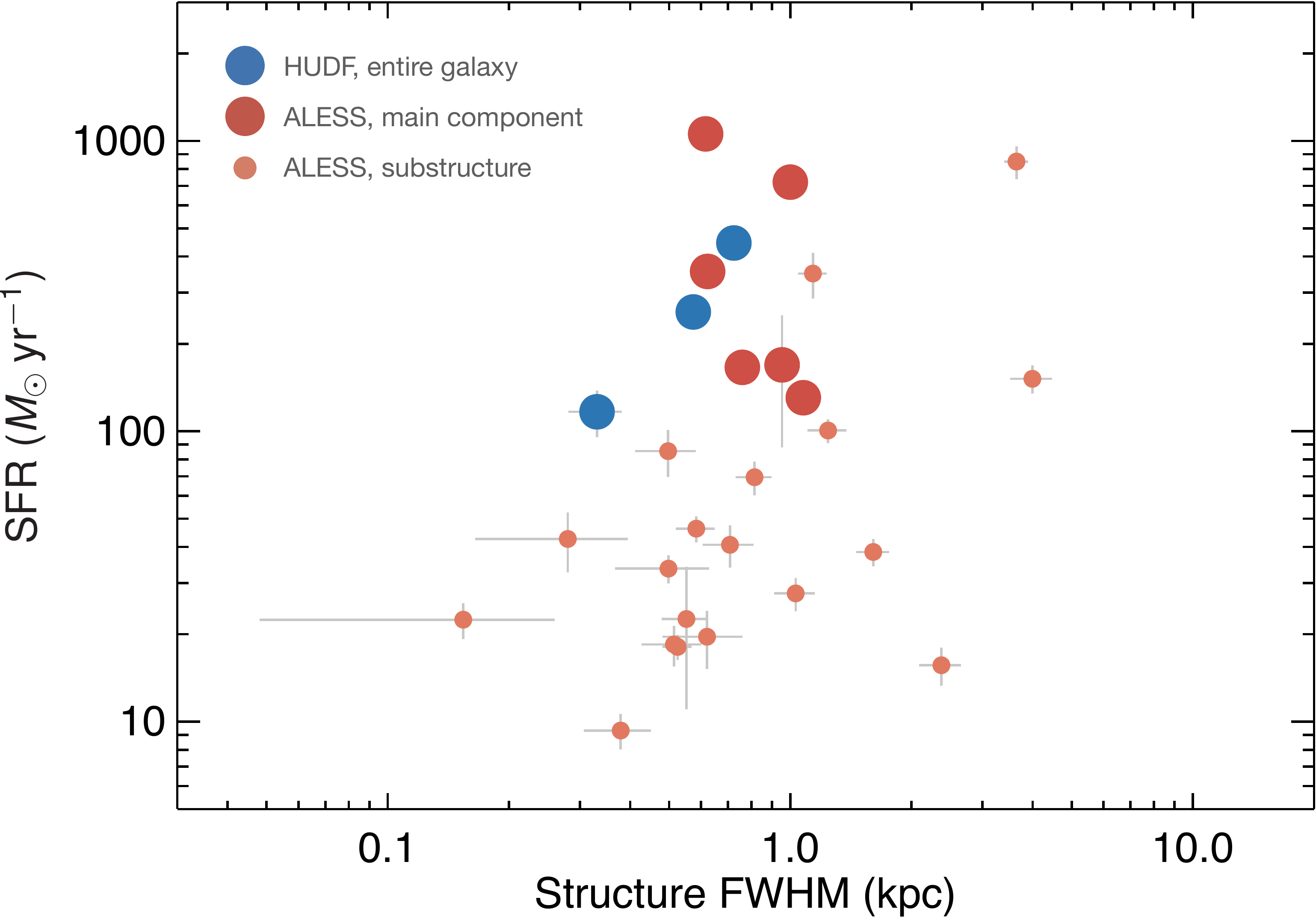}}
\caption{As shown in Figures \ref{fig:GILDAS_HUDF} and \ref{fig:GILDAS_ALESS}, HUDF galaxies only have smooth components of dust emission, whereas ALESS galaxies have multiple off-center substructures in addition to the main components. Some of the ALESS substructures have comparable sizes and SFRs to an entire HUDF galaxy. \label{fig:compsize}}
\end{figure}

The fact that the H19 substructures are found to be as bright (or in some cases brighter) than the entire HUDF galaxies, illustrated in Figure \ref{fig:compsize}, has an important observational implication. Specifically, some of the H19 substructures have comparable S$_{870 \mu m}$ and physical sizes to our faintest HUDF galaxy, UDF7, e.g., structure\# 3 in ALESS 3.1 and structure\# 3 in ALESS 17.1, among others (Table B1). Since UDF7 is significantly detected  ($\gtrsim 17 \sigma$),  even near the edge of our primary beam, the substructures such as those in the H19 SMGs would also be well detected if they are present in the HUDF galaxies. This negates the concern that the twice higher resolution of our observations (i.e., four times smaller beam area) may lack the surface brightness sensitivity to detect substructures such as those of H19.

\begin{figure*}
\figurenum{6}
\centerline{\includegraphics[width=\textwidth]{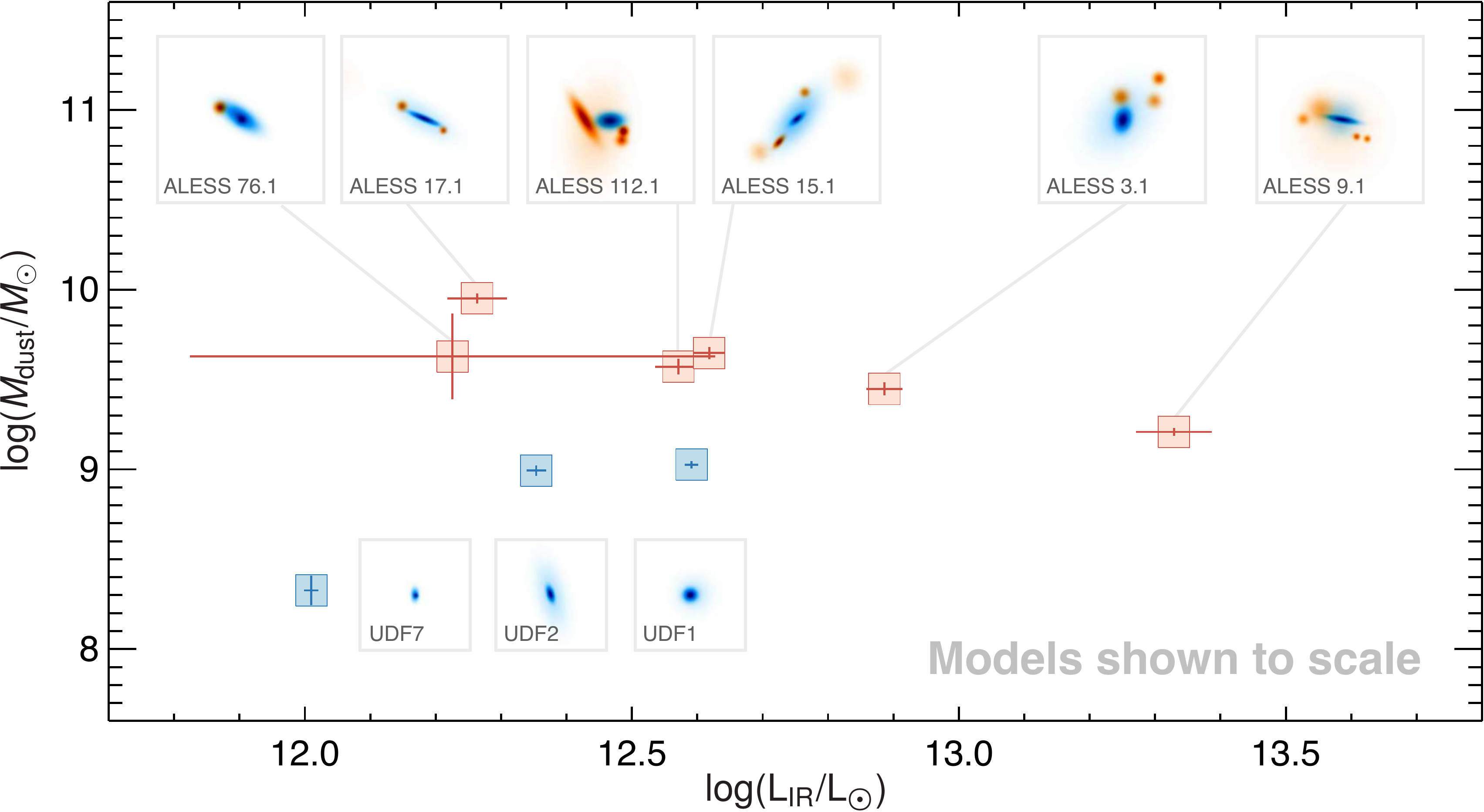}}
\caption{Our HUDF galaxies are all clumpless and H19 ALESS SMGs are clumpy with some substructures as large and/or as bright as our HUDF galaxies. A natural explanation is that H19 SMGs are interacting systems involving multiple objects similar to our HUDF galaxies. Furthermore, the significantly larger dust masses of the H19 SMGs could suggest exceptionally intense and/or recent starbursts (Section \ref{sec:discuss_SFG_SMG}). Models are shown to the same angular scale; same color coding as in Figures \ref{fig:GILDAS_HUDF} and \ref{fig:GILDAS_ALESS}. \label{fig:mdust_vs_LIR}}
\end{figure*}

\subsubsection{How are typical SFGs and SMGs different?}\label{sec:discuss_SFG_SMG}

To take a closer look at the physical differences (and commonalities) between our HUDF galaxies and the H19 SMGs, we estimate the $L_{\rm IR}$, SFR, and stellar masses of the ALESS sample consistently using the same method as described in Section \ref{sec:obs_ancdata} using the most recent photometry and redshift. The most significant revisions in the available data from the previously published estimates are the {\it HST} $H_{160}$ photometry from \citet{Chen15} and the 2-mm photometry from ALMA program\# 2015.1.00948.S (PI: da Cunha). The optical-to-8 \micron\ and $24-870$ \micron\ photometry were taken from \citet{Simpson14} and \citet{Swinbank14}, respectively, with the addition of the VIDEO and HSC observations \citep{Jarvis13, Ni19}; the $z_{\rm spec}$ were from H19.

While the $M_{\rm dust}$ and $L_{\rm IR}$ are relatively well constrained with the far-IR photometry, estimating the stellar mass is challenging for the H19 SMGs. They are undetected or not significantly detected at any optical/IR bands except at $3.6-8.0$ \micron. As a result, the star-formation history (SFH), age, metallicity, and extinction, on which stellar mass estimates strongly depend, cannot be firmly constrained. To quantify the extent of these systematics, we experimented with two independent methods, deriving the ALESS stellar masses using the {\tt HYPERZ} \citep{Bolzonella00} and the \citet{PPG08} codes (i.e., the same as with the HUDF galaxies in Section \ref{sec:obs_ancdata}), and compared them with the H19 estimates based on {\tt MAGPHYS}. For the {\tt HYPERZ} fit, we adopt a constant SFH, the BC03 models with stellar population age priors ranging from $0.05-0.4$ Gyr, Chabrier IMF, and the Calzetti attenuation law. For the estimates with the \citet{PPG08} code, we adopt exponentially declining BC03 models with Chabrier IMF, the Calzetti attenuation law and no assumptions about metallicity. Additionally, we explore a range of $\tau$ from a single stellar population to constant star formation models; $A_V$ up to 5 mag is allowed. We find a broad agreement of stellar mass estimates from the two methods, but they are significantly lower (in two cases an order of magnitude lower) than the estimates from {\tt MAGPHYS}. While some of these differences can be attributed to the assumptions of stellar population properties, both {\tt MAGPHYS}' and our approaches could be affected by different pitfalls: ours from not having a firm constraint on dust extinction (whereby $A_V$ is underestimated), and {\tt MAGPHYS} from assuming the energy balancing argument despite the dislocations between the stellar mass build-up and intensely star-forming regions (illustrated by the optical/submillimeter comparisons in Figure 2 of H19). As a result, {\tt MAGPHYS} could systematically overestimate $A_V$ of any stellar mass build-up that is not co-spatial with the dusty star-forming regions. It is possible that our stellar mass estimates are too low and H19's too high. The conundrums facing both the conventional and energy-balancing approaches will require {\it JWST} and next-generation extremely large telescopes to definitively resolve \citep[see also a comprehensive analysis of various stellar mass estimators  of SMGs based on simulations by][]{Michalowski14}. For the purpose of comparing the H19 and HUDF samples, we adopt the average of the estimates from our two methods as H19 stellar masses, tabulated in Table \ref{tab:ALESS_sourcetable}, along with H19 values for comparison\footnote{For ALESS9.1, the IRAC photometry is blended with a nearby bright object, e.g., Figure 10 of \citet{Chen15}. We estimated the stellar mass with deblended photometry out to 4.5 \micron}. At the broadest level, this exercise cautions that care is needed in interpreting results pertaining to stellar masses for these extremely dust-obscured SMGs.

Even with a rudimentary constraint on stellar masses, it emerges that four in six H19 SMGs have comparable stellar masses to our HUDF galaxies. These four SMGs are within the scatter of the main sequence (Figure \ref{fig:mainseq}), rendering them ``typical SFGs'' by our definition. They also have similar $L_{\rm IR}$, which could imply similar SFRs barring the probability of top-heavy IMFs at higher \SFRSD\ (discussed in Section \ref{sec:results_SFRSD}). The H19 SMGs appear to have larger dust masses, as evident in the $3-5\times$ brighter far-IR fluxes (Figure \ref{fig:mdust_vs_LIR}). However, this might not be representative of typical submillimeter-selected galaxies because H19 galaxies are among the brightest at the far-IR from the \citet{Hodge16} sample, which in-turn are among the brightest from the \citet{Hodge13} sample. Also, their larger dust masses do not necessarily imply larger gas mass, because a lower GDR might be more appropriate if they are merger-driven starbursts. This is because strong star-bursting systems typically have a CO-to-$M_{\rm gas}$ $\alpha_{\rm CO}$ conversion factor of 0.8 M$_{\odot}$/(K km s$^{-1}$ pc$^{2}$), which corresponds to GDR $\approx 30$ \citep[e.g.,][]{Leroy11, Magdis12, Genzel12, Silverman18}. It is, therefore, possible that both the ALESS and HUDF samples have comparable gas masses.

Given the comparable $L_{\rm IR}$, $M_*$, and perhaps even $M_{\rm gas}$, between the four out of six H19 SMGs and our HUDF galaxies, why are H19 SMGs much more dust-rich and clumpy? In other words, four of H19 SMGs and our HUDF galaxies are physically similar, so why do they appear differently? An explanation for the H19 substructures being comparably bright to our HUDF galaxies might be that these substructures originated from a class of disruptive event involving multiple systems with dust/gas reservoirs at the scale of our HUDF galaxies, e.g., major mergers of typical SFGs. The scale of the disruptive event found in our HUDF galaxies, characterized by the lack of star-forming substructures at our resolution and sensitivity, could be less violent, e.g., gas-rich disk instability or minor mergers. We note that at higher redshifts, minor mergers have been proposed as disruptive events triggering intense star formation in submillimeter-bright galaxies \citep[e.g.,][at $z \sim 4.5$]{GomezGuijarro18}. Our interpretation of the H19 substructures diverges from that put forward by Hodge et al. that the SMG substructures are potentially signatures of the spiral, ring, and bar structures induced by interactions. Since such structures were not observed in the less violent interactions in our HUDF galaxies, we feel that it is unlikely that the more disruptive interactions characteristic of SMGs will lead to the formation of galactic structures akin to those found in the local Universe. Rest-frame optical images could help resolve the question of the degree of disturbance of the galaxies and whether they are possible merger products. The {\it HST} $H_{160}$ imaging of the HUDF galaxies reaches a sensitivity of 29.8 mag AB \citep{Illingworth13}, but that of the ALESS galaxies has a median sensitivity of only 27.8 mag AB \citep{Chen15}. More sensitive near-IR images are needed to systematically characterize the morphologies of H19 SMGs.

The picture of H19 SMGs representing a more extreme class of disruptive event that induces exceptionally intense starbursts (even though not all of them are presently starbursting) might also explain their larger dust mass. A merger-driven starburst in a gas-rich environment could proceed very rapidly compared to its gas re-accretion timescale (e.g., $0.05-0.1$ Gyr). As such, dust and metal contents build up quickly in the observed star-forming regions with rapid consumption of gas, leading to a lower GDR even before an appreciable stellar mass increase occurs \citep[see, e.g., the enhanced $M_{\rm dust}/M_*$ in starburst galaxies reported by][]{Bethermin15}. We stress that the proposed scenario is based on a very small number of galaxies. A larger sample of spatially-resolved molecular gas distributions as well as \SFRSD\ maps from multiple bands to accurately constrain the $T_{\rm dust}$ for a large diversity of SFGs will be vital to testing these possible explanations and is already feasible with ALMA.

\subsubsection{Clumplessness vs. models predicting clumps}\label{sec:discuss_model}

Our observations of three typical SFGs at  $z \sim 3$, selected largely on the basis of stellar mass, shows their star formation to be smoothly distributed without significant clumps. This contrasts with the H19 SMGs, selected on the basis of brightness at 870 \micron, which finds significant clumpiness in galaxies of similar stellar mass and far-IR luminosity. However, even in these cases, the clumps account for a minority of the far-IR luminosity and hence most of the star formation is not clumpy in both samples. Likewise, the \citet{Genzel11} images of $z \simeq 2.3$ galaxies in H$\alpha$, whose stellar masses and SFRs are similar to our HUDF SFGs, indicate some clumps (Section \ref{sec:discuss_field}) but, again, accounting for only a small fraction of the total SFR. Studies of lensed galaxies (Section \ref{sec:discuss_lensed}) also show no convincing cases where the star formation is largely due to clumps.

When observers started to identify giant clumps at $z \sim 2$ using {\it HST} images and found them to be $\sim 1$ kpc in size and $\sim 10^9$ \Msun\ in stellar mass, theorists invoked gas fragmentation driven by gravitational instability in gas-rich turbulent disks fed by intense inflows of gas fuel to reproduce such clumps in simulations of both isolated galaxies  \citep{Noguchi99, Immeli04, Bournaud07} and in cosmological simulations \citep{Agertz09, Ceverino10, Genel12}. As observational evidence is mounting that the apparent giant clumps are likely caused by the limited angular resolution of observing facilities and the modest S/N of the observations, and any clumps are intrinsically an order of magnitude smaller (Section \ref{sec:discuss_lensed}), the spatial and/or temporal  resolution  of  models have also improved by an order of magnitude along with the more realistic sub-grid physics, affording vast improvement in simulation of disk substructures. With the improved models, a new consensus emerged among modelers that typical clumps formed by disk fragmentation should be an order of magnitude smaller than previously thought, with clump mass in the range of 10$^7 - 10^8$ \Msun\ and clump radii of $100-200$ pc \citep{Moody14, Mandelker14, Mayer16, Oklopcic17, Mandelker17}, and that clumps more massive than $10^9$ \Msun\ are rare, likely the result of clump-clump mergers or further gas accretion \citep{Tamburello15, Tamburello17}.

To reconcile this prediction of smaller clumps with the observations of kpc-sized clumps, \citet{Behrendt16} showed that clusters of $\sim10$ small clumps, each of $\sim10^7$ \Msun\ and $\sim70$ pc in diameter, can appear as a single giant clump when observed at the resolution of {\it HST} and AO-assisted ground-based facilities. Perhaps more importantly, these clump clusters can explain the large velocity dispersions observed in the giant clumps  \citep[$50-100$ km\,s$^{-1}$,][]{Genzel11} as due to the internal motion within the clump  cluster,  i.e., without having to invoke stellar feedback. \citet{Tamburello17} constructed mock H$\alpha$ maps from the simulations of \citet{Tamburello15} and ``observed'' them at 0.1 and 1 kpc resolutions. They found that the inferred physical properties of clumps depend sensitively on the observing resolution, e.g., the clump masses differ by a factor of two and the typical clump radii by an order of magnitude between the 0.1 and 1 kpc observations. 

More recently, Faure et al. (in preparation) have used hydrodynamic simulations with a sub-pc resolution to produce mock observations at ``{\it HST}-like'' and ``ALMA-like'' resolutions, 800 and 200 pc, respectively. At the {\it HST}-like resolution, the gas disk morphologies are dominated by giant clumps each containing a few percent (up to $5-10$\%) of the galaxy's gas mass; simulations suggest that these giant gas clumps are the counterparts of optical giant clumps. When observed at the 200-pc resolution of our ALMA observations, these giant clumps break into several smaller clouds. These clouds, resolved in ALMA images, typically contain $0.2-0.5$\% of the total gas mass of their host galaxy, with an upper limit of 1.5\%. This picture can be tested with ALMA observations slightly more sensitive than those of our HUDF galaxies on a larger sample. These simulations agree with \citet{DessaugesZavadsky17}, \citet{Fisher17}, and \citet{Rigby17} that the blending of multiple small star-forming clumps can explain the apparent clumpiness at the kpc-scale. Faure et al. (in preparation) also show that these apparent giant kpc-scale clumps, containing multiple smaller clouds, are not transient chance superpositions but gravitationally bound structures that can be gradually dispersed by stellar feedback and/or migrate toward the galactic bulge.

In addition to ruling out the presence of kpc-scale giant star-forming clumps, our observations are in tension with some of the recent models. Since the intrinsic radii of clumps from simulations and observations of gravitational lens systems are now known to be $50-200$ pc, our 200-pc observations are well-suited for detecting them. For example, the typical in-situ clumps predicted by \citet{Mandelker17} are $10^{7.5} - 10^{8.5}$ \Msun\ in mass, $300-800$ pc in diameter, and harbor SFR $\sim0.1-10$\% of that in the disk. The brighter populations of these predicted clumps should be well detected in our galaxies, contrary to our results. Models predicting lower mass clumps are still consistent with the smooth appearance of our HUDF galaxies. However, the finding that the Eyelash is clumpless (Section \ref{sec:discuss_lensed}) will pose an even stronger tension on clumps, down  to a physical scale of $50-100$ pc. That is, while some of these simulations are in agreement with the rest-frame UV clumps revealed by gravitational lenses, the situation remains unconstrained in typical massive main-sequence SFGs at $z \simeq 3$. 

We note that the clumps in the formative era of the Milky Way that \citet{Clarke19} propose to explain the present-day $\alpha$-abundance bimodality measured by the SDSS/APOGEE survey \citep{Nidever14} are $10^{7.5} - 10^{9.5}$ \Msun\ in mass; the predicted distribution peaks at 10$^{8.1}$ \Msun. The low-mass side of this range is still consistent with our HUDF galaxies being smooth.

\section{CONCLUSION}\label{sec:conclusion}

We present 200-pc resolution ALMA continuum images at rest-frame $\simeq 240$ \micron\ of three typical SFGs at $z \sim 3$ and compare them with those of six SMGs from \citet{Hodge19} observations with comparable image fidelity, whose morphological properties have been derived using a self-consistent procedure. Our results can be summarized as follow.

\begin{itemize}

\item Our images trace cold dust, which reveals the central obscured star-forming regions in typical SFGs to have smooth, disk-like morphology $\simeq 1-3$ kpc across. Our images are sensitive to SFR of $1-3$ \Msunyr\ ($1\sigma$) at a 200-pc scale. Any clumps or substructures can contain no more than $1-7$\% ($3\sigma$ upper limits) of the total star formation.

\item No other peripheral dust substructure is seen outside the intense star-forming region in these three typical SFGs. Two of our HUDF galaxies have in total $\approx 10$ UV-selected star-forming clumps. These clumps are $\simeq 2-10$ kpc from the intense star-forming region. No enhancement of the dust emission is observed from these UV-selected clumps.

\item The absence of dust substructures at the 200-pc scale supports the picture that the apparent kpc-sized star-forming clumps are results of clusters of $\lesssim100$ pc sub-clumps blending together due to the $\simeq 0\farcs1$ resolution of optical observations. However, the brighter populations of the model-predicted sub-clumps should already be detectable in our ALMA observations, contrary to our findings.

\item In contrast with our HUDF galaxies, SMGs in the \citet{Hodge19} sample have multiple dust substructures, with individual substructures containing typically $10-30$\% of the total SFR (considerably larger than $2-8$\% found by Hodge et al., see discussion in Section \ref{sec:discuss_H19}), and being as large as our HUDF galaxies in some cases. A natural explanation is that these SMGs are interacting systems involving multiple objects at the scale of our HUDF SFGs. Nonetheless, clumps account for only a minority of the star formation even in these cases.

\end{itemize}

Additional spatially-resolved ALMA observations of cold gas kinematics will be required to confirm that the disk-like cold gas distributions in typical SFGs are indeed rotationally supported, as well as characterizing the potentially strong outflows, which will be critical to interpreting the roles of these compact, intense central star-forming regions in the formation of the bulge. Our findings that typical SFGs are smooth while SMGs are clumpy are still based on a very small number of galaxies. Significant ALMA time investment will be necessary to construct representative samples of typical SFGs and SMGs with high-fidelity morphological information.

\section*{Acknowledgments}
We thank the anonymous referee for helpful suggestions and comments. The authors would like to thank Jacqueline Hodge, Rob Ivison, Erica Keller, Victor de Souza Magalhaes, J\'erome Pety, Gerg\"{o} Popping, Alvio Renzini, David Ruffolo, Kamolnate Trisupatsilp, and Fabian Walter for helpful discussions and suggestions. W.R. acknowledges support from the Thailand Research Fund/Office of the Higher Education Commission Grant Number MRG6280259; Chulalongkorn University's CUniverse and the Ratchadapiseksompot Endowment Fund. P.G.P-G. acknowledges support from the Spanish Government grant AYA2015-63650-P. G.E.M. acknowledges support from the Villum Fonden research grant 13160 ``Gas to stars, stars to dust: tracing star formation across cosmic time'', the Cosmic Dawn Center of Excellence funded by the Danish National Research Foundation and  the ERC Consolidator Grant funding scheme (project ConTExt, grant number No. 648179). K.K. and T.W. acknowledge the support from JSPS KAKENHI Grant Number JP17H06130 and the NAOJ ALMA Scientific Research Grant Number 2017-06B. C.C.W. acknowledges support from the NSF Astronomy and Astrophysics Fellowship grant AST-1701546. Kavli IPMU is supported by World Premier International Research Center Initiative (WPI), MEXT, Japan. This  paper makes use  of  the  following ALMA data: ADS/JAO.ALMA\#2017.1.00001.S, \#2016.1.00048.S, \#2015.1.00948.S, \#2012.1.00173.S. ALMA is a partnership of ESO (representing its member states), NSF (USA) and NINS (Japan), together with NRC (Canada) and NSC and ASIAA (Taiwan)  and KASI (Republic  of Korea), in cooperation with the Republic of Chile. The Joint ALMA Observatory is operated by ESO, AUI/NRAO and NAOJ.

\begin{appendix}

\section{GALFIT Modeling of {\it HST} $H_{160}$ Images}

\begin{figure*}[h]
\figurenum{1}
\centerline{\includegraphics[width=\textwidth]{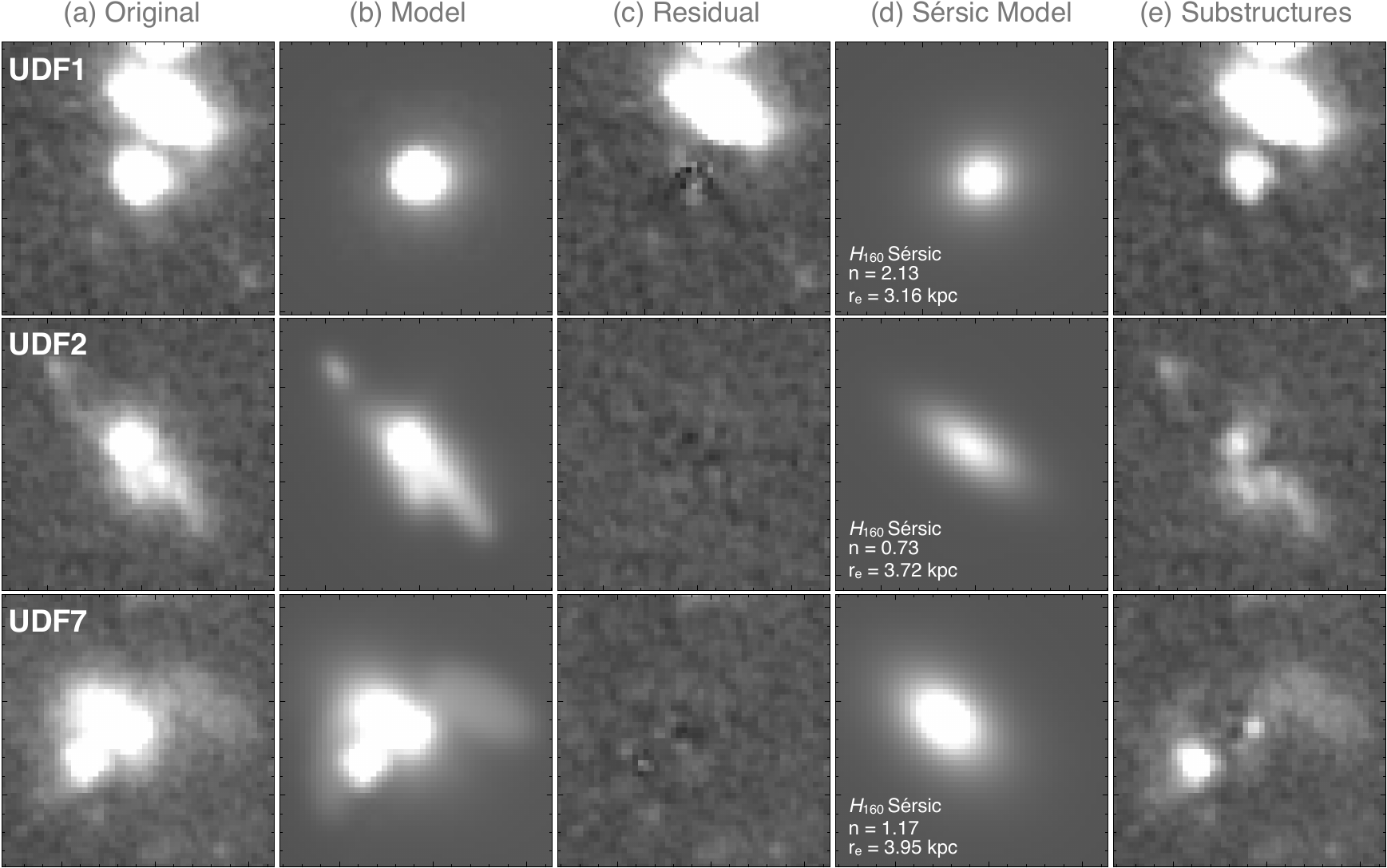}}
\caption{Steps taken to search for stellar mass substructures in the {\it HST} $H_{160}$ observations, which probe the rest-frame optical emission. Briefly, we model all emission components (column b) to achieve a uniform residual (column c), then only subtract the dominant S\'ersic model (column d) to produce substructure maps (column e). All images are shown with an identical linear scale. \label{fig:A1}}
\end{figure*}

To search for stellar mass substructures, we model the {\it HST}/WFC3 $H_{160}$ images with GALFIT \citep{Peng10} using the procedure illustrated in Figure A\ref{fig:A1}. Starting with the original image in column (a), we model the emission with a combination of multiple S\'ersic and/or point sources as necessary to achieve a uniform residual. The best-fit model and the resulting residual are shown in columns (b) and (c), respectively. From these best-fit models, we subtract only the dominant S\'ersic components, shown in column (d), to produce the stellar mass substructure maps in column (e). These are the maps shown in the third column from the left of Figure A\ref{fig:compare_HST_ALMA}, along with the S\'ersic index $n$ and effective radius $r_e$ overlaid in the corresponding cutout. 

\section{Properties of substructures in the Hodge et al. (2019) SMGs}

Results of $uv$-plane source decomposition for the \citet{Hodge19} SMGs are tabulated in Table B1 (cf. Table 3 for our HUDF sample). These source components are visualized in Figure 3 (cf. Figure 2 for our HUDF sample). The $uv$-based analysis is described in Section 4.2.3.

\begin{deluxetable*}{lcccccc}
\tablenum{1}
\tablecaption{H19 source decomposition in the $uv$ plane \label{tab:Comp_ALESS}}
\tablehead{
\colhead{ID} & \colhead{Component} & \colhead{Type} & \colhead{RA} & \colhead{Dec} & \colhead{Flux} & \colhead{Size}\\
\colhead{} & \colhead{} & \colhead{} & \colhead{($''$)} & \colhead{($''$)} & \colhead{($\mu$Jy)} & \colhead{(mas)}
}
\startdata
ALESS 3.1 &  Total &          & 	 03:33:21.514          &  $-$27:55:20.466   	   & $ 9134 \pm  308 $  &\\
		 & 1 & Main	 & $   \phantom{-}0.014 \pm 0.010 $  & $  \phantom{-}0.009 \pm 0.011 $  & $ 4646 \pm  248 $  & $  541 \pm  29 \times  379 \pm  18 $ \\
		 & 2 & Main	 & $  -0.006 \pm 0.001 $  & $ -0.006 \pm 0.002 $  & $ 2860 \pm  131 $  & $  150 \pm   5 \times  103 \pm   4 $ \\
		 & 3 & Clump	 & $   \phantom{-}0.011 \pm 0.005 $  & $ \phantom{-}0.199 \pm 0.005 $  & $  723 \pm   93 $  & $  109 \pm  11\phantom{0} $\\
		 & 4 & Clump	 & $  -0.329 \pm 0.005 $  & $  \phantom{-}0.365 \pm 0.004 $  & $  481 \pm   47 $  & $   78 \pm   8\phantom{0} $\\
		 & 5 & Clump	 & $  -0.289 \pm 0.007 $  & $  \phantom{-}0.164 \pm 0.006 $  & $  423 \pm   69 $  & $   95 \pm  13 $\\
ALESS 9.1 &  Total &          & 	 03:32:11.332	       &    $-$27:52:12.005     & $ 9767 \pm  590 $  &\\
		 & 1 & Main	 & $  -0.022 \pm 0.005 $  & $ -0.004 \pm 0.001 $  & $ 1535 \pm  139 $  & $  194 \pm  11 \times   39 \pm   5\phantom{00} $ \\
		 & 2 & Main	 & $  -0.009 \pm 0.011 $  & $  \phantom{-}0.003 \pm 0.006 $  & $ 2710 \pm  350 $  & $  226 \pm  18\phantom{0} $\\
		 & 3 & Clump	 & $   \phantom{-}0.177 \pm 0.010 $  & $  \phantom{-}0.085 \pm 0.007 $  & $ 1403 \pm  235 $  & $  178 \pm  14\phantom{0} $\\
		 & 4 & Clump	 & $   \phantom{-}0.017 \pm 0.020 $  & $ -0.101 \pm 0.025 $  & $ 3409 \pm  381 $  & $  572 \pm  39\phantom{0} $\\
		 & 5 & Clump	 & $  -0.243 \pm 0.008 $  & $ -0.177 \pm 0.006 $  & $  171 \pm   38 $  & $   44 \pm  17 $\\
		 & 6 & Clump	 & $   \phantom{-}0.332 \pm 0.007 $  & $ -0.002 \pm 0.006 $  & $  343 \pm   59 $  & $   78 \pm  13 $\\
		 & 7 & Clump	 & $  -0.149 \pm 0.004 $  & $ -0.158 \pm 0.003 $  & $  194 \pm   32 $  & $    ... $ \\
ALESS 15.1 & Total  &          & 	03:33:33.371           &    $-$27:59:29.720      & $ 9471 \pm  252 $  &\\
		   & 1 & Main	 & $   \phantom{-}0.001 \pm 0.004 $  & $  \phantom{-}0.003 \pm 0.005 $  & $ 6204 \pm  187 $  & $  \phantom{0}440 \pm   5 \times  201 \pm  17 $ \\
		   & 2 & Main	 & $  -0.002 \pm 0.003 $  & $  \phantom{-}0.003 \pm 0.002 $  & $  900 \pm   89 $  & $  101 \pm   6 \times   40 \pm   8\phantom{0} $ \\
		   & 3 & Clump	 & $   \phantom{-}0.162 \pm 0.003 $  & $ -0.205 \pm 0.003 $  & $  673 \pm   72 $  & $   \phantom{0}95 \pm   9 \times   41 \pm  10 $\\\   
		   & 4 & Clump	 & $  -0.446 \pm 0.011 $  & $  \phantom{-}0.372 \pm 0.009 $  & $  768 \pm   81 $  & $  202 \pm  18\phantom{0} $\\
		   & 5 & Clump	 & $  -0.072 \pm 0.005 $  & $  \phantom{-}0.242 \pm 0.004 $  & $  369 \pm   58 $  & $   64 \pm  10 $\\
		   & 6 & Clump	 & $   \phantom{-}0.337 \pm 0.009 $  & $ -0.295 \pm 0.007 $  & $  553 \pm   71 $  & $  129 \pm  14 \phantom{0}$\\
ALESS 17.1 & Total  &          & 	03:32:07.285           &    $-$27:51:20.892	   & $ 9133 \pm  322 $  &\\
		   & 1 & Main	 & $   \phantom{-}0.024 \pm 0.008 $  & $  \phantom{-}0.015 \pm 0.005 $  & $ 4129 \pm  235 $  & $  455 \pm  17 \times  174 \pm   9 $ \\
		   & 2 & Main	 & $   \phantom{-}0.007 \pm 0.002 $  & $  \phantom{-}0.005 \pm 0.001 $  & $ 3121 \pm  178 $  & $  191 \pm   5 \times   36 \pm   2 $ \\
		   & 3 & Clump	 & $   \phantom{-}0.206 \pm 0.002 $  & $  \phantom{-}0.118 \pm 0.002 $  & $  790 \pm   67 $  & $   62 \pm   5 $\\
		   & 4 & Clump	 & $  -0.166 \pm 0.003 $  & $ -0.102 \pm 0.003 $  & $  407 \pm   53 $  & $   44 \pm   8 $\\
		   & 5 & Clump	 & $   \phantom{-}0.772 \pm 0.019 $  & $  \phantom{-}0.361 \pm 0.018 $  & $  683 \pm   96 $  & $  280 \pm  33 $\\
ALESS 76.1 &  Total &           & 	03:33:32.350           &   $-$27:59:55.735   & $ 4428 \pm  158 $  &\\
		   & 1 & Main	 & $   \phantom{-}0.002 \pm 0.004 $  & $  \phantom{-}0.019 \pm 0.003 $  & $ 3594 \pm  117 $  & $  271 \pm  13 \times  117 \pm   4 $ \\
		   & 2 & Main	 & $  -0.008 \pm 0.003 $  & $ -0.008 \pm 0.002 $  & $  313 \pm   48 $  & $   \phantom{0} ... $\\
		   & 3 & Clump	 & $   \phantom{-}0.185 \pm 0.005 $  & $  \phantom{-}0.104 \pm 0.004 $  & $  521 \pm   93 $  & $   74 \pm   9 $\\
ALESS 112.1 & Total  &         & 03:32:48.856 	       &    $-$27:31:13.192      & $ 6436 \pm  301 $  &\\
		    & 1 & Main	 & $   \phantom{-}0.007 \pm 0.003 $  & $ -0.014 \pm 0.002 $  & $ 1976 \pm  100 $  & $  \phantom{0}170 \pm   8 \times  101 \pm   5 $ \\
		    & 2 & Clump	 & $   \phantom{-}0.242 \pm 0.005 $  & $  \phantom{-}0.007 \pm 0.007 $  & $ 1524 \pm  131 $  & $  290 \pm  17 \times   78 \pm   8\phantom{0} $ \\
		    & 3 & Clump	 & $  -0.097 \pm 0.007 $  & $ -0.189 \pm 0.009 $  & $  297 \pm   66 $  & $   \phantom{0}75 \pm  17 $\\
		    & 4 & Clump	 & $   \phantom{-}0.132 \pm 0.020 $  & $ -0.100 \pm 0.022 $  & $ 2296 \pm  238 $  & $  \phantom{0}627 \pm  63 \times  378 \pm  41 $ \\
		    & 5 & Clump	 & $  -0.113 \pm 0.003 $  & $ -0.108 \pm 0.002 $  & $  340 \pm   46 $  & $   \phantom{0}18 \pm  12 $
\enddata
\tablecomments{Source IDs are from \citet{Hodge19}. RA and Dec of each component are tabulated as offsets in arcseconds relative to the phase center position (ICRS) in the corresponding `Total' row. Sizes are FWHM; dots in the size column indicate an unresolved component. Fluxes are integrated, observed-frame 870 \micron\ flux.}
\end{deluxetable*}

\end{appendix}

\end{document}